\documentclass[final,authoryear,1p,times]{elsarticle}

\usepackage{amsmath,amssymb,xspace,subfigure}
\usepackage{amsfonts}
\usepackage[usenames,dvipsnames]{xcolor}
\usepackage[]{graphicx}
\usepackage{url}
\usepackage{hyperref}
\usepackage{fullpage}
\usepackage{url}
\usepackage{float}
\usepackage{bm}
\usepackage{soul}
\usepackage{xcolor}
\usepackage{ulem}
\newcommand{\fold}{\mathrm{fold}} 
 
\newcommand{\mudot}{\dot{\mu}}

\graphicspath{{./Figures/}}

\journal{Elsevier}

\begin{document}

\begin{frontmatter}



%\title{Delayed bifurcation in elastic ribbons: path control and mode selection}

\title{Exploiting dynamic bifurcation in elastic ribbons for mode skipping and selection}

% \red{how about this title} \blue{It is good. I would prefer to combine skipping and selection, like "for mode skipping and selection"}

% Exploiting delayed bifurcation in elastic ribbons to selectively skip modes


\author[a]{Weicheng Huang}
\author[b]{Tian Yu}
\author[c]{Dominic Vella}
\author[d,e]{K Jimmy Hsia\corref{cor5}}
\author[d,f]{Mingchao Liu\corref{cor5}}
\cortext[cor5]{Corresponding authors: kjhsia@ntu.edu.sg (K.J.H); m.liu.2@bham.ac.uk (M.L.)}

\address[a]{School of Engineering, Newcastle University, Newcastle upon Tyne NE1 7RU, UK}
\address[b]{Department of Mechanics and Aerospace Engineering, Southern University of Science and Technology, Shenzhen, 518055, China}
\address[c]{Mathematical Institute, University of Oxford, Oxford OX2 6GG, UK}
\address[d]{School of Mechanical and Aerospace Engineering, Nanyang Technological University, Singapore 639798, Republic of Singapore}
\address[e]{School of Chemistry, Chemical Engineering and Biotechnology, Nanyang Technological University, \\ Singapore 639798, Republic of Singapore}
\address[f]{Department of Mechanical Engineering, University of Birmingham, Birmingham B15 2TT, UK}

\begin{abstract}

In this paper, we systematically study the dynamic snap-through behavior of a pre-deformed elastic ribbon by combining theoretical analysis, discrete numerical simulations, and experiments.
By rotating one of its clamped ends with controlled angular speed, we observe two snap-through transition paths among the multiple stable configurations of a ribbon in three-dimensional (3D) space, which is different from the classical snap-through of a two-dimensional (2D) bistable beam.
Our theoretical model for the static bifurcation analysis is derived based on the Kirchhoff equations, and dynamical numerical simulations are conducted using the Discrete Elastic Rods (DER) algorithm. The planar beam model is also employed for the asymptotic analysis of dynamic snap-through behaviors.
The results show that, since the snap-through processes of both planar beams and 3D ribbons are governed by the saddle-node bifurcation, the same scaling law for the delay applies.
We further demonstrate that, in elastic ribbons, by controlling the rotating velocity at the end, distinct snap-through pathways can be realized by selectively skipping specific modes, moreover, particular final modes can be strategically achieved.
Through a parametric study using numerical simulations, we construct general phase diagrams for both mode skipping and selection of snapping ribbons.
The work serves as a benchmark for future investigations on dynamic snap-through of thin elastic structures and provides guidelines for the novel design of intelligent mechanical systems.

\end{abstract}


\end{frontmatter}

% \linenumbers

\section{Introduction}
\label{sec:Intro}

Slender structures exist in both natural and industrial settings, taking on various forms such as blood vessels, human hair, plant branches, thin films, and airplane wings. Characterized by their large slenderness ratio, i.e., the cross-sectional dimensions are much smaller compared to their axial length \citep{simone2007introduction}, slender structures are often modeled as either one-dimensional (1D) elements like beams and rods (length $L$ $\gg$ width $W$ $\sim$ thickness $h$) or two-dimensional (2D) elements such as plates and shells ($L$ $\sim$ $W$ $\gg$ $h$). Ribbons, with three distinct scales (i.e., $L$ $\gg$ $W$ $\gg$ $h$), are also a type of slender structure. Slender structures often exhibit geometric non-linearities even if the material properties remain linear, leading to structural instabilities including buckling, snapping, and folding, under moderate external forces or simple boundary conditions \citep{audoly2010elasticity,reis2018mechanics}. These instabilities normally result in dramatic shape changes and nonlinear mechanical responses \citep{radisson2023designing}, benefiting the design of advanced metamaterials and intelligent systems \citep{hu2015buckling, bertoldi2017flexible, faris2007mechanical}. Applications of such instability phenomena range from soft robotics \citep{qin2023modeling,yang2023morphing} to flexible electronics \citep{bo2023mechanically} and biomedical devices \citep{zhang2019multifunctional}. Thus, understanding the mechanics of slender structures is essential for uncovering pattern-formation mechanisms in nature and developing innovative devices with advanced functionalities.

Among various nonlinear behaviors of slender structures, snap-through buckling represents a unique instability phenomenon, where an elastic object suddenly `jumps' from one equilibrium state to another, accompanied by a sudden release of elastic energy \citep{pandey2014dynamics,gomez2017critical}. One typical example of this rapid transition observed in everyday experiences is the inversion of umbrellas in strong winds. At the same time, nature also harnesses this effective mechanism to engineer biological systems, such as Venus flytraps, which exploit this phenomenon to quickly release energy that has been stored, thus enabling them to catch prey \citep{forterre2005venus}. Taking inspiration from the natural world, a number of engineering applications have sought to make use of snap-through instabilities of slender structures including, but not limited to, shape-morphing structures with tunable surface topographies \citep{holmes2007snapping} and geometry configurations \citep{liu2023snap,yang2023morphing}, snapping metamaterials with multi-stabilities caused by periodic arrangement of snapping units \citep{lu2023multiple2,lu2023multiple1,rafsanjani2016snapping,mao2022modular,rodriguez2023mechanical}, soft actuators with application to soft robotics \citep{overvelde2015amplifying,chen2018harnessing,wang2023insect}, and fluidic devices for flow control at both macro- \citep{gomez2017passive} and nano-scales \citep{jiao2020snap}.

Rational designs of engineering systems utilizing snap-through buckling depend on a deep understanding of the mechanics of snap-through phenomena. While extensive existing researches have focused on identifying the onset of snap-through in typical structures like beams, plates, and shells under varying loads or boundary conditions \citep{camescasse2013bistable,zhang2020configurations,ma2022snap,keleshteri2018snap,qiao2020elastic,wan2023finding}, the dynamics of snap-through remains less studied. Previous studies have substantially overestimated snap-through speed by considering the balance of elastic and inertial forces. For instance, the timescale of leaf closure in Venus flytraps is theoretically estimated to be around $0.001$ – $0.01$ s, whereas the actual observed motion takes approximately $0.1$ – $1$ s \citep{forterre2005venus}. This discrepancy is often attributed to the intrinsic energy dissipation mechanisms of the material, such as poroelasticity or viscoelasticity. Such hypotheses have been particularly relevant in cases involving fluid-filled porous materials, where the internal fluid flow takes time \citep{liu2019multiscale}, or the silicone-based elastomers, where viscoelasticity significantly influences the dynamic behaviors, leading to phenomena like pseudo-bistability \citep{brinkmeyer2012pseudo,gomez2019dynamics,chen2023pseudo} or acquired bistability \citep{urbach2020predicting}.

However, recent studies have demonstrated the occurrence of anomalous delayed snap-through in elastic structures in both experiments and simulations, even in the absence of dissipation mechanisms \citep{gomez2017critical,sano2018snap}, in which the snap-through transition is slowed significantly by the dynamic effect. This phenomenon is commonly referred to as the \emph{delayed instability}. The intrinsic nature of this delay in the snap-through transition can be understood in terms of a saddle-node (fold) bifurcation. In this scenario, as the control parameter changes, the current equilibrium state of the system abruptly ceases to exist. This transition is critically influenced by the proximity to the bifurcation point, where the net force acting on the structure is very small since the various forces within the structure are balanced (as it represents an equilibrium solution). Consequently, the movement of the structure slows down significantly \citep{gomez2017critical,liu2021delayed}. Such delayed instability has been demonstrated in an elastic beam by considering the effect of the proximity to bifurcation on the duration of snap-through for the case with fixed boundary, which is slightly away from the fold point \citep{gomez2017critical} and further extended by \cite{liu2021delayed} to consider the linear ramping of the bifurcation parameter. The scaling laws for the delay of both cases are obtained through asymptotic analysis and validated by numerical and experimental results.

It is noted that most of the previous works on snap-through dynamics are restricted to 2D circumstances and, in particular, any twist along the centerline of the strip is ignored. Recently, the snapping and bifurcation of pre-deformed ribbons in 3D space under stretching~\citep{morigaki2016stretching}, twisting~\citep{sano2019twist}, shearing~\citep{yu2019bifurcations,huang2020shear}, and indentation~\citep{huang2021snap} have begun to be studied. However, these studies mainly focus on the onset and the corresponding static features. It has been shown that the stability and bifurcation behaviors of ribbons in 3D space are much more complex compared to those of beams in 2D space. In particular, by controlling the uniaxial compression and transverse shear applied to a simple ribbon with clamping boundaries, four different stable configurations, referred to as different modes, exist and several different transition paths connecting different modes can be achieved under different loading/boundary conditions \citep{huang2020shear,huang2021snap}. While the static snap-through buckling behavior of elastic ribbons has been studied, the dynamics associated with this phenomenon are still largely unexplored. The investigation on the dynamic snap-through of elastic beams has indicated that the mode transition in multistable structures can be regulated through applying dynamic boundary conditions \citep{liu2021delayed,radisson2023designing,radisson2023dynamic}.

In this work, we focus on investigating the dynamical behaviors of snap-through buckling of narrow elastic ribbons (i.e., $L$ $\gg$ $W$ $\gg$ $h$) through theoretical analysis, numerical simulations, and simple tabletop experiments. Such understanding can be used to manipulate the transition between different modes as well as to select particular modes. To capture the static bifurcation behaviors of narrow ribbons, we employ the anisotropic Kirchhoff rod theory \citep{audoly2010elasticity}, and the corresponding ordinary differential equations (ODEs) are solved by using the continuation package AUTO 07P \citep{doedel2007lecture}. To investigate the dynamic snap-through buckling of ribbons, we adopt the numerical simulation method based on discrete differential geometry (DDG), in particular, the Discrete Elastic Rods (DER) algorithm -- a numerical approach to simulate 1D filament that has shown high computational efficiency in the field of computer graphics as well as mechanical engineering \citep{bergou2008discrete,bergou2010discrete}. In addition, we perform simple tabletop experiments to validate the model predictions of the configuration and transition between different modes. Finally, we numerically demonstrate the tunability of the snap-through paths and modes by changing the dynamical loading conditions.

This paper is organized as follows. Section 2 introduces the model setup. Section 3 reviews the standard Kirchhoff rod theory and the discrete numerical framework. The simplified planar beam for asymptotic analysis is followed up in Section 4. Section 5 and Section 6 delve into the static and dynamic snapping bifurcation analysis of elastic ribbons, respectively. Section 7 discusses the concepts of mode skipping and selection. Finally, concluding remarks are presented in Section 8.

\section{Problem setup}
\label{sec:problemsetup}

We consider a narrow elastic ribbon subjected to both uniaxial compression (i.e.~the end-shortening) and transverse shear to form the buckled configuration in 3D space. We then rotate one end of the buckled ribbon with constant angular speed to trigger the dynamic snap-through instability.

\begin{figure}[h!]
\centering
\includegraphics[width=1.00\textwidth]{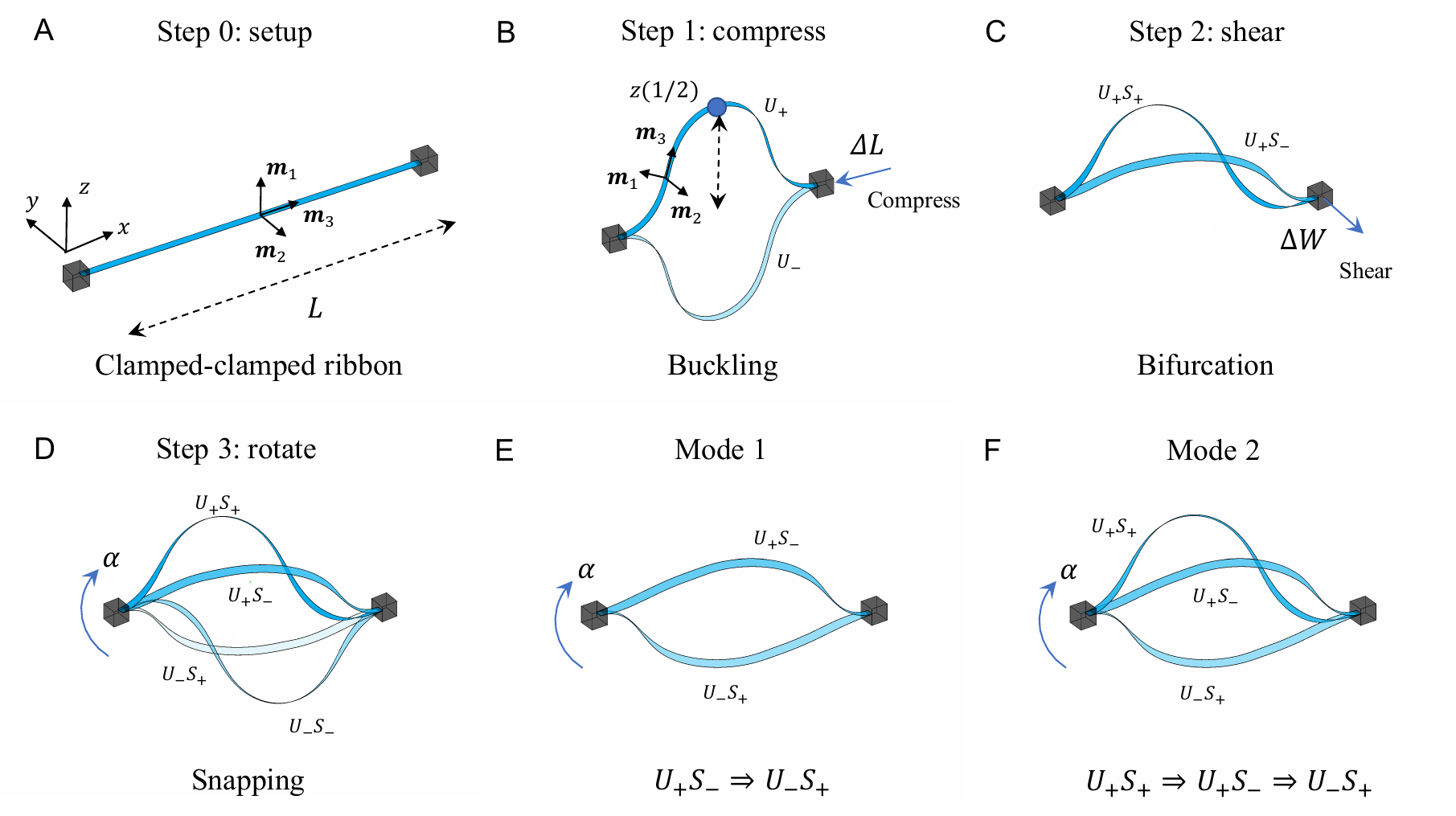}
\caption{Problem setup. (A) A naturally straight elastic ribbon with length $L$ (in $x$ direction), width $W$ (in $y$ direction), and thickness $h$ (in $z$ direction). (B) The ribbon undergoes out-of-plane buckling to form mode $U_+$ (or $U_-$) when subjected to compression $\Delta L$ in the $x$ direction. (C) The buckled ribbon undergoes supercritical pitchfork bifurcation to form mode $U_+S_-$ or $U_+S_+$ by further applying shear $\Delta W$ to the right end in $y$ direction. (D) The snap-through buckling of the bifurcated ribbon can be triggered by rotating one end with angle $\alpha$ to achieve the mode transition through two possible paths: (E) Path $1$: $U_+S_- \rightarrow U_-S_+$ and, (F) Path $2$: $U_+S_+ \rightarrow U_+S_- \rightarrow U_-S_+$. (The nomenclatures are based on the midpoint orientation  of the ribbon.)}
\label{fig:setupPlot}
\end{figure}

Figure~\ref{fig:setupPlot}A shows a naturally straight ribbon with length $L$, width $W$, and thickness $h$. In Step 1, we compress the ribbon along $x$-axis to trigger the out-of-plane buckling (with compression $\Delta L$) into a stable `$U$'--like shape,  donated as `$U_{+}$' for $z>0$ or `$U_{-}$' for $z<0$ (see Fig.~\ref{fig:setupPlot}B). In Step 2, a transverse shear $\Delta W$ is applied to one end along the $y$-direction. It is found that the ribbon exhibits a supercritical pitchfork bifurcation to form two possible equilibrium configurations (mode `$U_{+}S_{+}$' and `$U_{+}S_{-}$', see Fig.~\ref{fig:setupPlot}C) depending on the shear displacement \citep{yu2019bifurcations}. Note that there are also two mirror-symmetric equilibrium states, i.e., `$U_{-}S_{+}$' mode and `$U_{-}S_{-}$' mode, see Fig.~\ref{fig:setupPlot}D (not shown in Fig.~\ref{fig:setupPlot}C). Among them, some modes are symmetric (e.g., $U_{+}S_{+}$ and $U_{-}S_{-}$) and some are reverse-symmetric (e.g., $U_{+}S_{+}$ and $U_{+}S_{-}$). The structural symmetry is discussed in {\color{blue}Appendix A}, and the bifurcation phase diagram for different patterns is provided in {\color{blue}Appendix B}. In step 3 (see Fig.~\ref{fig:setupPlot}D), we dynamically rotate one end of the ribbon with an angular speed $\dot{\alpha}$. The ribbon will transition from one mode to another via snap-through buckling. Depending on its initial mode upon applying the shear displacement, there are two transition paths: Path 1: $U_{+}S_{-} \rightarrow U_{-}S_{+}$ (Fig.~\ref{fig:setupPlot}E), and Path 2: $U_{+}S_{+} \rightarrow U_{+}S_{-} \rightarrow U_{-}S_{+}$ (Fig.~\ref{fig:setupPlot}F).

\begin{figure}[t!]
\centering
\includegraphics[width=0.90\textwidth]{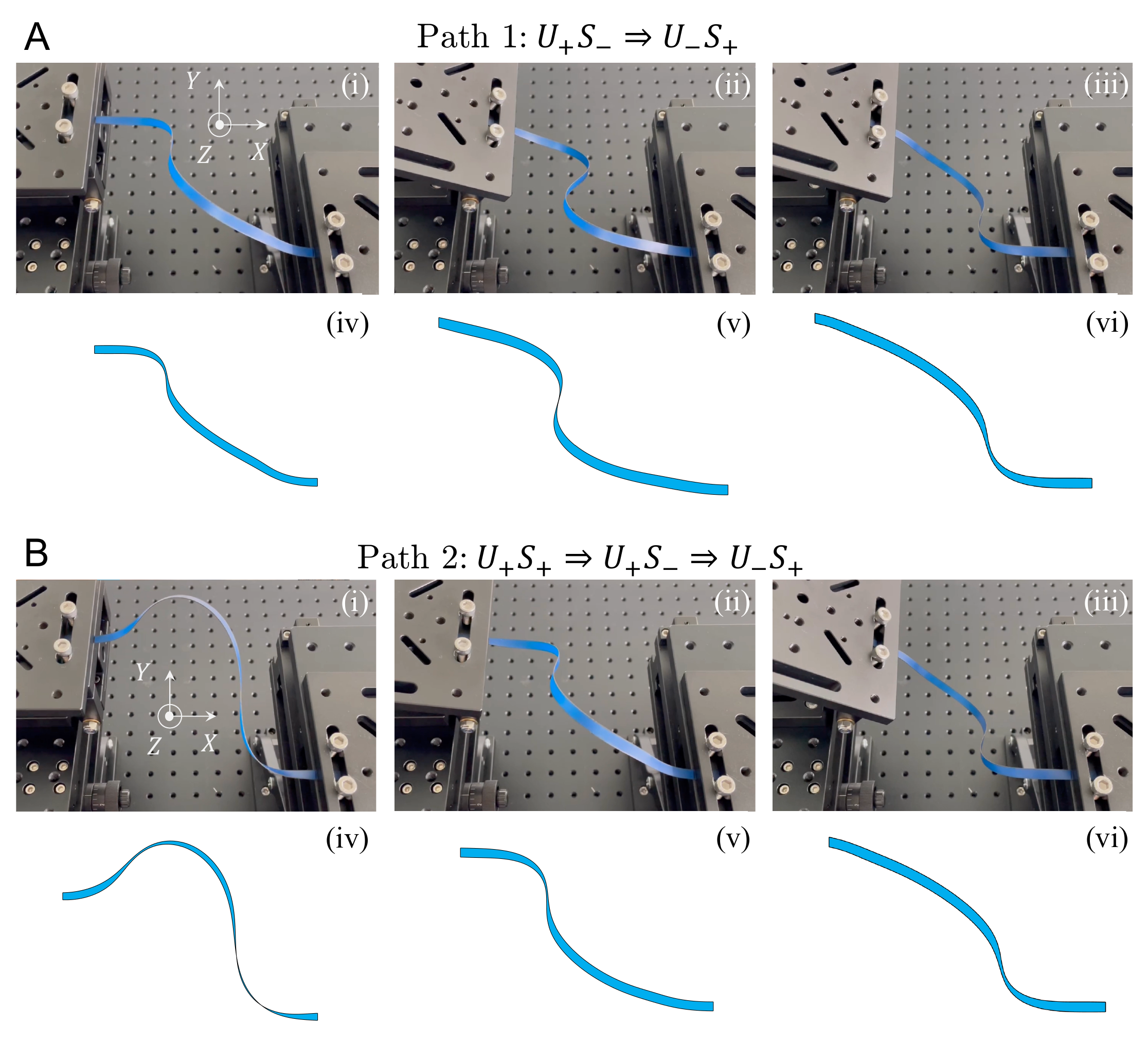}
\caption{Snapshots of the elastic ribbon (top view) during the snap-through transition process for: (A) Path $1$: $U_{+}S_{-} \rightarrow U_{-}S_{+}$ and, (B) Path $2$: $U_{+}S_{+} \rightarrow U_{+}S_{-} \rightarrow U_{-}S_{+}$, which include (i)-(iii) Experimental configurations and, (iv)-(vi) the corresponding simulated configurations. 
}
\label{fig:expSimPlot}
\end{figure}

As shown in Fig.~\ref{fig:expSimPlot}, the side-by-side comparisons between the tabletop experiments (i-iii) and the numerical simulation results (iv-vi) for those paths are presented.
Videos for the two snap-through paths from both tabletop experiments and numerical predictions are available in {\color{blue}Appendix C}.
Note that, as we rotate the left end, two mirror-symmetric paths will be triggered: (i) $U_{+}S_{+} \rightarrow U_{-}S_{-}$ and, (ii) $U_{-}S_{+} \rightarrow U_{+}S_{+} \rightarrow U_{-}S_{-}$. In the following sections, we employ the rod theory, numerical simulations, and experiments to investigate the dynamics of the rotation-induced snap-through of the ribbon, and further explore the skipping and selection of different modes of the ribbon under different dynamic loading rates.

\section{3D rod model}
\label{sec:3drodmodel}

In this section, we first review the Kirchhoff rod model for thin and narrow ribbons and then introduce the discrete numerical model based on DER method.

\subsection{Anisotropic rod model}
\label{sec:3dribbontheory}

We employ the Kirchhoff rod model to study the nonlinear mechanics of narrow ribbons  \citep{antman19755,nizette1999towards,ameline2017classifications}.
The deformed configurations of 1D rod-like structures can be accurately captured by Kirchhoff rod theory  ~\citep{goyal2005nonlinear, yu2019bifurcations, sano2019twist}. In this study, we consider a typical ribbon with $L \gg W $ and $W \sim O(10h)$, such that the anisotropic rod model is sufficiently accurate to perform the geometrically nonlinear mechanics in a narrow ribbon~\citep{yu2019bifurcations, huang2020shear,audoly2021one}. 
The Kirchhoff rod model is summarized as follows.

\paragraph{Kinematic} Consider a thin elastic ribbon of length $L$, width $W$ and 
thickness $h$, and made of isotropic and linearly elastic material with Young's modulus $E$, shear modulus $G$, and density $\rho$.
The ribbon can be described by its centerline $\mathbf{r}(s,t)$ and an associated right-handed orthonormal material frame $\{ \mathbf{m}_{1} (s,t), \mathbf{m}_{2} (s,t), \mathbf{m}_{3} (s,t) \}$, aligned to the width, the surface normal, and the local tangent of the ribbon's centerline, respectively. Here, $s \in [0,L]$ corresponds to the arc length of the undeformed centerline and will be used as a curvilinear coordinate to describe the dynamics of the ribbon. Throughout the paper, $(\;)'$ and $\dot{(\;)}$ are used to represent the derivatives with respect to spatial coordinate $s$ and time $t$, respectively. We have $\mathbf{m}_{3}  = \mathbf{r}' / || \mathbf{r}' ||$
and the local stretch of the centerline can be given as
\begin{equation}
\epsilon  = || \mathbf{r}'  || - 1,
\label{eq:continousStretching}
\end{equation}
The rotation of the material frame along the centerline can be formulated as
\begin{equation}
\begin{aligned}
\mathbf{m}'_{1} &= \bm{\omega}  \times \mathbf{m}_{1}, \\
\mathbf{m}'_{2} &= \bm{\omega}  \times \mathbf{m}_{2}, \\
\mathbf{m}'_{3} &= \bm{\omega}  \times \mathbf{m}_{3}, 
\end{aligned}
\end{equation}
where $\bm{\omega}$ corresponds to the Darboux vector,
\begin{equation}
\bm{\omega}   = \kappa_{1}  \mathbf{m}_{1}  + \kappa_{2}  \mathbf{m}_{2}  + \tau  \mathbf{m}_{3}  .
\label{eq:continousCurvatures}
\end{equation}
Here, $\kappa_{1} $ and $\kappa_{2} $ are the bending curvatures and $\tau $ is the twisting curvature.
The macroscopic strains for the 1D rod include the axial stretch in Eq.(\ref{eq:continousStretching}) and the curvatures in Eq.(\ref{eq:continousCurvatures}).

\paragraph{Constitutive law} A linear constitutive law between strains and forces/moments is given as, 
\begin{equation}
\begin{aligned}
\mathbf{N}  &= EA \epsilon  \mathbf{m}_{3}, \\
\mathbf{M}  &= EI_{1} \kappa_1  \mathbf{m}_{1}  + EI_{2}  \kappa_2   \mathbf{m}_{2}   + GJ \tau   \mathbf{m}_{3}  .
\end{aligned}
\end{equation}
where $EA$ is the stretching stiffness, $EI_1$ and $EI_2$ represent the two bending rigidities, and $GJ$ corresponds to the torsional stiffness.
For a rod with a rectangular cross-section, the stretching, bending, and twisting stiffness are,
\begin{equation}\label{eq:inertiaofmoment} 
EA = E Wh, \quad EI_1=\frac{1}{12} E W^3 h, \quad EI_2=\frac{1}{12} E W h^3, \quad
GJ=\lambda G  W h^3.
\end{equation}
where $\lambda$ depends on the cross-section of the rod.
It is noted that the stretching of the rod's centerline is more expensive than the bending and twisting because of its special geometry, e.g., $L \gg W$, so that the stretching is almost zero everywhere and bending and twisting deformations dominate, i.e., the rod is inextensible and unshareable, $\mathbf{N}= \mathbf{0}$.

\paragraph{Equilibrium equations} The dynamic equilibrium equations are the statement of force and moment balance, 
\begin{equation}
\begin{aligned}
\rho A \ddot{\mathbf{r}} &= \mathbf{N}' + \mathbf{f}_{e}, \\
\mathbf{\dot{\bm{\Omega}}} & = \mathbf{M}' + \mathbf{m}_{3} \times \mathbf{N}' + \bm{\tau}_{e}.
\end{aligned}
\label{eq:EquilibriumEquations}
\end{equation}
where $\mathbf{f}_{e}$ is the external force density per unit length, $\bm{\tau}_{e}$ is the external torque density per unit length, and $\bm{\Omega}$ is the angular momentum density per unit length.

\paragraph{Boundary conditions} For a clamped-rotate strip subject to uniaxial compression, $\Delta L$, transverse shear, $\Delta W$, and the rotational speed of one end $\dot{\alpha}$, the associated boundary conditions are,
\begin{equation}
\begin{aligned}
\mathbf{r}(0, t) &= [0, 0, 0], \\
\mathbf{m}_{1}(0, t) &= [ \sin (\alpha_0 +\dot{\alpha} t) , 0, \cos (\alpha_0 +\dot{\alpha} t) ], \\
\mathbf{m}_{2}(0, t) &= [ 0, -1, 0 ], \\
\mathbf{m}_{3}(0, t) &= [ \cos (\alpha_0 +\dot{\alpha} t) , 0, -\sin (\alpha_0 +\dot{\alpha} t) ], \\
\mathbf{r}(L, t) &= [ {L - \Delta L}, -{\Delta W}, 0], \\
\mathbf{m}_{1}(L, t) &= [ 0, 0, 1 ], \\
\mathbf{m}_{2}(L, t) &= [ 0, -1, 0 ], \\
\mathbf{m}_{3}(L, t) &= [ 1, 0, 0 ]. \\
\end{aligned}
\label{eq:boundaryConditions}
\end{equation}
Continuation package AUTO 07P ~\citep{doedel2007lecture} is used to solve the static version of Eqs.(\ref{eq:EquilibriumEquations}) and (\ref{eq:boundaryConditions}) (i.e., $\rho A \ddot{\mathbf{r}}$, $\mathbf{\dot{\bm{\Omega}}}$ and $\dot{\alpha} t$ are omitted) with $\alpha_0$ treated as a continuation parameter and material frames at the two ends implemented through quaternions.   
AUTO solves the nonlinear boundary value problem through the collocation method and keeps tracking bifurcations and limit points, and is able to switch onto the bifurcated branches ~\citep{doedel2007lecture}.

\subsection{Discrete numerical model}
\label{sec:discreteModel}

Here, we adopt the well-established DER method to simulate the nonlinear dynamic behaviors of an elastic ribbon \citep{huang2020shear,huang2021snap}.
In the DER method, the centerline of the rod is discretized into $ N $ nodes and $ N - 1 $ edges.
To follow the convention of DER, we use subscripts to denote quantities associated with the nodes, e.g., $\mathbf x_i$ for position of the $i$-th node, and superscripts when associated with edges, e.g., $\mathbf e^i$ for the $i$-th edge.
A bar on top indicates the evaluation of the undeformed configuration, e.g., $ || \bar{\mathbf{e}}^{i} || $ is the undeformed length of $i$-th edge.
Each edge has an orthonormal adapted reference frame $ \left\{\mathbf{d}^{i}_{1}, \mathbf{d}^{i}_{2}, \mathbf{d}^{i}_{3} \right\} $ and an orthonormal material frame $ \left\{\mathbf{m}^{i}_{1}, \mathbf{m}^{i}_{2}, \mathbf{m}^{i}_{3}\right\} $; both frames share the tangent as one of the directors, 
\begin{equation}
\mathbf{d}^{i}_{3} \equiv \mathbf{m}^{i}_{3} = \mathbf e^i / || \mathbf e^i ||.
\end{equation}
The rotational angle between the reference frame and material frame is denoted as $ \theta^{i} $, i.e.,
\begin{equation}
\begin{aligned}
\mathbf{m}_{1}^{i} &= \mathbf{d}_{1}^{i} \cos \theta^{i} + \mathbf{d}_{2}^{i} \sin \theta^{i}, \\
\mathbf{m}_{2}^{i} &= \mathbf{d}_{2}^{i} \cos \theta^{i} - \mathbf{d}_{1}^{i} \sin \theta^{i}.
\end{aligned}
\end{equation}
Nodal positions together with twist angles constitute $ 4N-1 $ degrees of freedom (DOF) vector of a single rod system, 
\begin{equation}
\mathbf{q} = \left[\mathbf{x}_{0}, \theta^{0}, \mathbf{x}_{1}, ..., \mathbf{x}_{N-2}, \theta^{N-2}, \mathbf{x}_{N-1} \right].
\end{equation}
An elastic rod is modeled as a mass-spring system, with a lumped mass (and angular mass) at each node (and edge), and associated discrete stretching, bending, and twisting energies.
The discrete elastic stretching, bending, and twisting energies are given by~\cite{bergou2008discrete,bergou2010discrete}
\begin{equation}
\begin{aligned}
E_s &= \sum_{i=0}^{N-2} \frac {1} {2} EA (\epsilon^{i})^2 || \bar{\mathbf{e}}^{i} ||, \\
E_b &= \sum_{i=0}^{N-1}  \frac {1} {2}{EI}_{1}(\kappa_{1,i})^2 \Delta {l}_{i} + \sum_{i=0}^{N-1} \frac {1} {2} {EI}_{2}(\kappa_{2,i})^2 \Delta {l}_{i}, \\
E_t &= \sum_{i=0}^{N-1}  \frac {1} {2} 
 {GJ} ( \tau_{i} )^2 \Delta {l}_{i} ,
\end{aligned}
\label{eq:totalEnergy}
\end{equation}
where $ \epsilon^{i} $ is the discrete stretching strain associated with the $i$-th edge,
$ \kappa_{1,i} $ and $ \kappa_{2,i} $ are the discrete bending curvatures at the $i$-th node,
$ \tau_{i} $ is the twist at the $i$-th node, 
and $ \Delta {l}_{i} = \left( || \bar{\mathbf{e}}^{i} || + || \bar{\mathbf{e}}^{i+1} || \right) / 2 $ is the Voronoi length at $i$-th vertex.
As the structure considered here is naturally straight, the undeformed curvatures are zeros, e.g., $\bar{\kappa}_{1, i} = 0$.
The microscope strains, e.g., $ \epsilon^{i} $, $ \kappa_{1,i} $, $ \kappa_{2,i} $, and $ \tau_{i} $, can be expressed in terms of the $11$ consecutive DOFs near $i$-th node, $\{ \mathbf x_{i-1}, \theta^{i-1}, \mathbf x_{i}, \theta^{i}, \mathbf x_{i+1} \}$,
\begin{equation}
\begin{aligned}
\epsilon^i &= \frac{ || \mathbf e^i ||}{|| \bar{\mathbf{e}}^i ||} -1, \\
\kappa_{1,i} & = \frac { \mathbf{e}^{i-1} \times \mathbf{e}^{i} } { || \mathbf{e}^{i-1} || \cdot || \mathbf{e}^{i} || + \mathbf{e}^{i-1} \cdot \mathbf{e}^{i} } \cdot \frac {\left( \mathbf m_2^{i-1} + \mathbf m_2^i \right)} {\Delta l_{i}}, \\
\kappa_{2,i} & = - \frac {\mathbf{e}^{i-1} \times \mathbf{e}^{i}} { || \mathbf{e}^{i-1} || \cdot || \mathbf{e}^{i} || + \mathbf{e}^{i-1} \cdot \mathbf{e}^{i} } \cdot \frac{\left( \mathbf m_1^{i-1} + \mathbf m_1^i \right) } {\Delta l_{i}}, \\
\tau_{i} &= \frac { \left( \theta^{i} - \theta^{i-1} + {m}_{i}^{\mathrm{ref}} \right) }{\Delta l_{i}},
\end{aligned}
\label{eq:strainEquation}
\end{equation}
where $ {m}_{i}^{\mathrm{ref}} $ is the reference twist associated with the reference frame~\citep{bergou2008discrete, bergou2010discrete, jawed2018primer}.
The elastic forces (associated with nodal positions) and elastic moments (associated with the twist angles) are the negative gradients of total potentials,
\begin{equation}
\mathbf{F}^{\text{int}} = - \frac { \partial } {\partial \mathbf{q}} \left( E_s+E_b+E_t \right).
\label{elasticForceRod}
\end{equation}
Finally, the implicit Euler method is adopted to update the DOF vector ${\mathbf {q}}$ from time step $ t_{k} $ to $ t_{k+1} = t_{k} + \delta t$ (in which $ \delta t $ is the time step size):
\begin{equation}
\begin{aligned}
{\mathbb{M}} \mathbf{\ddot{q}}(t_{k+1}) &- {\mathbf{F}}^{\text{dam}}(t_{k+1}) -  {\mathbf{F}}^{\text{int}}(t_{k+1})  = \mathbf{0}, \\ 
&{\mathbf{q}}(t_{k+1}) = {\mathbf{q}}(t_{k}) + \dot{\mathbf{q}}(t_{k+1})  \delta t, \\
& \dot{\mathbf{q}}(t_{k+1}) = \dot{\mathbf{q}}(t_{k}) + \mathbf{\ddot{q}}(t_{k+1})  \delta t,
\end{aligned}
\label{eq:NewmarkBeta}
\end{equation}
where $\mathbb{M}$ is the time-invariant mass matrix, and the damping force vector can be calculated as
\begin{equation}
{\mathbf{F}}^{\text{dam}}(t_{k+1}) = -\nu {\mathbb{M}} \dot{\mathbf{q}}(t_{k+1})
\label{Dampingforcevector}
\end{equation}
where $\nu$ is the dimensionless damping coefficient, which will be defined later.

\noindent
To achieve a specified the boundary conditions, in our numerical simulation, the first two nodes $\{ \mathbf{x}_{0}, \mathbf{x}_{1} \}$ and the last two nodes $\{ \mathbf{x}_{N-2}, \mathbf{x}_{N-1} \}$, as well as the first twisting angle $\theta_{0}$ and the last twisting angle $\theta_{N-2}$, are constrained based on the compression, shear, and rotation.
All other nodes and edges are free to evolve according to the dynamic equilibrium.

\section{Reduced planar beam model}
\label{sec:reducedbeammodel}

In this section, we reduce the 3D rod model presented in Section \ref{sec:3dribbontheory} into the planar beam theory for analysis. 
We first assume an inextensible planar rod and obtain the dynamic elastica equation; then consider the small deflection and small rotation scenario such that the theory converges to the classical Euler–Bernoulli beam model, which will be used for the asymptotic analysis.

\subsection{Planar elastica}
\label{sec:PlanarElasctia}

When the structure is planar and the associated transverse shear is zero (i.e., $\Delta W  = 0$), the ribbon model can be reduced to a beam model, where the rotations along the $\mathbf{m}_{1}$ and $\mathbf{m}_{3}$ directors are forbidden, resulting in
\begin{equation}
\kappa_{1}(s,t) \equiv 0 \; \mathrm{and} \;  \tau(s,t) \equiv 0.
\end{equation}
The deformed centerline can then be expressed as 
\begin{equation}
{\mathbf{r}}  = {x}  \mathbf{e}_{x} + {z}  \mathbf{e}_{z},
\end{equation}
and its target direction can be simplified as
\begin{equation}
{\mathbf{r}}'  = \cos \phi  \mathbf{e}_{x} + \sin \phi  \mathbf{e}_{z},
\end{equation}
where $\phi$ is the rotation of the beam centerline.
Further, the dynamic governing equation can be reduced to that of a planar beam as,
\begin{equation}
\rho I_{2} \ddot{\phi} = EI_{2} \phi'' - n_{x} \sin{\phi} + n_{z} \cos{\phi},
\label{eq:beamDyEq}
\end{equation}
where $n_{x}$ (and $n_{y}$) is the component of the resultant force attached to the centreline, 
\begin{equation}
\mathbf{n} = n_{x} \mathbf{e}_{x} + n_{z} \mathbf{e}_{z}.
\end{equation}
The associated boundary conditions are
\begin{equation}
\begin{aligned}
x(0, t) &= 0, \\
z(0, t) &=0, \\
\phi(0,t) &= \alpha_0  + \dot{\alpha} t, \\
x(L, t) &= {L - \Delta L}, \\
z(L, t) &=0, \\
\phi(L,t) &= 0.
\end{aligned}
\label{beamBoundaryConditions}
\end{equation}

\subsection{Quasi-linear approximation}
\label{sec:Quasilinearapproximation}

When the deflection of the beam is small, we can use $x$ rather than $s$ as the independent variable as the difference between the reference configuration and the deformed configuration is negligible.
Vertical displacement, $z(x,t)$, can be used to describe the deformation of an elastic beam. Eq. (\ref{eq:beamDyEq}) can be reduced to
\begin{equation}
\rho h \frac{\partial^2 z} {\partial t^2} + \Upsilon \frac{\partial z} {\partial t} + B \frac{\partial^4 z} {\partial x^4} + P_c \frac{\partial^2 z} {\partial x^2} =0,\ \ \ \ 0 < z < 1,
\label{nonscalebeamfun}
\end{equation}
where $B = EI_{2} / W$ is the bending stiffness, $P_{c}$ is the compressive force applied at the boundary, and the damping effect (per unit area of the beam and assumed constant for simplicity), $\Upsilon$, is  included in the equations.
The boundary conditions for the beam equation can be written as,
\begin{equation}
\begin{aligned}
z(0,t)&=0, \\ \frac{\partial z} {\partial x}(0,t)&= \alpha_0 + \dot{\alpha} t, \\ z(L,t)&=0, \\ \frac{\partial z} {\partial x}(L,t)&=0,
\end{aligned}
\label{nonscaleboundary}
\end{equation}

\noindent and the imposed end-shortening constraint is
\begin{equation}
\int^L_0\left(\frac{\partial z}{\partial x}\right)^2dx=2\Delta L.
\label{nonscaleconstraint}
\end{equation}\\
To non-dimensionalize the problem, we let
\begin{equation}
X = x/L,  \quad
Z = z / L \cdot (\Delta L / L)^{-1/2},  \quad
T =t/t^*,
\end{equation}
where $t^*$ is the inertial time scale 
\begin{equation}
t^* =\left(\frac{\rho h L^4}{B}\right)^{1/2}.
\end{equation}
\noindent
Introducing the scale parameters defined above, the beam equation Eq. \eqref{nonscalebeamfun} can be scaled as
\begin{equation}
\frac{\partial^2 Z}{\partial T^2} + \tilde{\nu} \frac{\partial Z}{\partial T} + \frac{\partial^4 Z}{\partial{X}^4} + {\sigma}^2 \frac{\partial^2 Z}{\partial{X}^2}=0,\ \ \ 0< X< 1,
\label{scaledbeamfun}
\end{equation}\\
where
\begin{equation}
{\sigma}^2=P_c \frac{L^2}{B},
\end{equation}
and
\begin{equation}
\tilde{\nu} = \frac {\Upsilon L^2}  {\sqrt{\rho h B}} = \nu \cdot t^*,
\end{equation}
where $\nu = {\Upsilon} / {\rho h}$, which is consistent with the dimensionless damping coefficient introduced in our discrete numerical simulations in Eq. \eqref{Dampingforcevector}.
\noindent
The boundary conditions can also be scaled as
\begin{equation}
\begin{aligned}
Z(0,T) &= 0, \\
Z(1,T) &= 0, \\
\frac{\partial Z} {\partial X} (0,T) &= \mu, \\
\frac{\partial Z} {\partial X} (1,T) &= 0, \\
\label{scaledboundary1}
\end{aligned}
\end{equation}\\
where
\begin{equation}
\mu = \mu_0 + \dot{\mu} T
\end{equation}
and 
\begin{equation}
\mu_0  =\alpha_0 \cdot \left({\Delta L} / {L} \right)^{-{1}/{2}}, \quad
\dot{\mu} = \dot{\alpha} \cdot t^* \cdot \left({\Delta L} / {L} \right)^{-{1}/{2}}.
\end{equation}

\noindent
Finally, the imposed end-shortening constraint becomes
%$$
%\int^L_0\left(\frac{\partial w}{\partial x}\right)^2dx = \int^\frac{L}{L}_0\frac{(L \Delta L)}{L^2}\left(\frac{\partial W}{\partial X}\right)^2d(XL)=2\Delta L,\\
%$$
%\noindent
%i.e.,
\begin{equation}
\int^1_0\left(\frac{\partial Z}{\partial X}\right)^2dX=2.
\label{scaledconstraint}
\end{equation}\\

\subsection{Static behavior of the quasi-linear problem}
\label{sec:Staticbehavioroflinearproblem}

The solution of the static beam equation Eq. \eqref{scaledbeamfun}  subjected to boundary conditions Eq. \eqref{scaledboundary1} can be obtained as\\
\begin{equation}
Z(X)=\frac{\sigma X(\cos\sigma-1)+\sigma [\cos\sigma (1-X)-\cos\sigma]-\sin\sigma X-\sin\sigma(1-X)+\sin\sigma}{\sigma(2\cos\sigma+\sigma \sin\sigma-2)}.
\label{equilibriumshapefun}
\end{equation}\\
To determine $\sigma$ in terms of the control parameter $\mu$, we substitute Eq. \eqref{equilibriumshapefun} into the end-shorting constraint Eq. \eqref{scaledconstraint}, ignore the time varying term $\dot{\mu} T$, and obtain
\begin{equation}
\mu^2 = \frac{8\sigma(2\cos\sigma+\sigma \sin\sigma-2)^2}{2{\sigma}^3-{\sigma}^2(\sin2\sigma+4\sin\sigma)+4\sigma(\cos\sigma-\cos2\sigma)+2(\sin2\sigma-2\sin\sigma)}.
\label{equilibriumconstraint}
\end{equation}\\
\noindent
For each value of $\mu$, the allowed values of $\sigma(\mu)$ can be calculated numerically using MATLAB. The bifurcation diagram can be expressed in terms of $Z(1/2)$ and $\sigma$ as
\begin{equation}
Z(1/2)=\mu \frac{\tan(\sigma/4)}{2\sigma},
\label{equilibriumconstraint12}
\end{equation}
which follows from Eq. \eqref{equilibriumshapefun}. This result shows that the saddle-node (fold) bifurcation at the critical state of
\begin{equation}
\mu_{\fold}= \pm 1.7818,\ \ Z_{\fold}(1/2) =  \mp 0.3476,\ \ \sigma_{\fold} = 7.5864.
\label{bifurcationsolution}
\end{equation}
These analytical solutions of the fold point will be validated by discrete numerical simulations in the next sections.

\subsection{Snap-through dynamics of the quasi-linear problem}
\label{sec:Snapthroughdynamicsoflinearproblem}

We rescale time as $T=\dot{\mu}^{-\eta}\mathcal{T}$ for some (currently unknown) exponent $\eta$; the beam equation Eq. \eqref{scaledbeamfun} then becomes\\
\begin{equation}
\dot{\mu}^{2\eta} \frac{\partial^2 Z}{\partial \mathcal{T}^2} + \tilde{\nu} \dot{\mu}^{\eta} \frac{\partial Z}{\partial \mathcal{T}} + \frac{\partial^4 Z}{\partial{X}^4} + {\sigma}^2 \frac{\partial^2 Z}{\partial{X}^2}=0,\ \ \ 0< X< 1,
\label{rescaledbeamfun}
\end{equation}\\

\noindent
The boundary condition Eq. \eqref{scaledboundary1} becomes
\begin{equation}
\begin{aligned}
Z(0,\mathcal{T}) &= 0, \\
Z(1,\mathcal{T}) & =0, \\
\frac{\partial Z}{\partial X}(0,\mathcal{T}) &= \mu(T) = \mu_0 +\dot{\mu}^{1-\eta}\mathcal{T}, \\
\frac{\partial Z}{\partial X} (1,\mathcal{T}) &= 0.
\label{rescaledboundary}
\end{aligned}
\end{equation}\\
The constraint condition Eq. \eqref{scaledconstraint} is
\begin{equation}
\int^1_0\left(\frac{\partial Z}{\partial X}\right)^2dX = 2.
\label{rescaledconstraint}
\end{equation}\\

\noindent
The asymptotic expansion about the fold shape can be chosen as
\begin{align}
Z(X,\mathcal{T})\ &=\ Z_{\fold}(X)+\mudot^{\gamma}\: Z_0(X,\mathcal{T})+\mudot^{2\gamma}\: Z_1(X,\mathcal{T})+\dots, \label{asymexpansionW}\\
\sigma(\mathcal{T})\ &=\ \sigma_{\fold}+\mudot^{\gamma}\: {\sigma}_0(\mathcal{T})+\mudot^{2\gamma}\: {\sigma}_1(\mathcal{T})+\dots.
\label{asymexpansionsigma}
\end{align} 

\noindent
Substituting Eq. \eqref{asymexpansionW} and Eq. \eqref{asymexpansionsigma} into Eq. \eqref{rescaledbeamfun}, we find that, since the fold shape $Z_\fold(X)$ is time-independent, the time derivatives only enter in the higher order terms. Consequently, the first non-trivial problem, which involves terms at $O(|\mudot|^{\gamma})$ in Eq. \eqref{rescaledbeamfun}, is quasi-static. We refer to this as the leading-order problem, since it involves the lowest-order perturbation $(Z_0,\sigma_0)$ to the fold shape in Eq. \eqref{asymexpansionW} and Eq. \eqref{asymexpansionsigma}. We will show below that this problem is, in fact, an eigenvalue problem for $(Z_0,\sigma_0)$ for which the time dependence only enters via the solution amplitude. It is only in the first-order problem, which occurs at $O(|\mudot|^{2\gamma})$ in Eq. \eqref{rescaledbeamfun}, that time-dependent terms appear. We therefore impose a balance between the inertia term in Eq. \eqref{rescaledbeamfun} and the first-order perturbation $(Z_1,\sigma_1)$, which gives, upon comparing exponents of $|\mudot|$, the relation $2\eta + \gamma = 2\gamma$. 

Moreover, to obtain non-trivial dynamics, the ramping term must enter the boundary condition at first-order, implying $2\gamma = 1-\eta$. Combining these relations gives
\begin{equation}
\eta = \frac{1}{5}, \quad \gamma = \frac{2}{5}.
\end{equation}
It is interesting to note that, although the boundary condition of the system in this work is different from that in a previous work \citep{liu2021delayed}, the basic dynamics remain the same. Following the  asymptotic analysis in \cite{liu2021delayed}, we can then obtain the scaling of the time delay for the underdamped case, where the inertia term dominates the dynamic behavior, as
\begin{equation}
T_s \sim \dot{\mu}^{-\frac{1}{5}} \sim \dot{\alpha}^{-\frac{1}{5}},
\label{timedelayunderdamp}
\end{equation}
and correspondingly, the angle delay is
\begin{equation}
\Delta \alpha = T_s \cdot \dot{\alpha} \sim\dot{\alpha}^{\frac{4}{5}}.
\label{angledelayunderdamp}
\end{equation}
However, for the overdamped case, the damping term (characterized by $\Lambda \sim \nu \mudot^{-1/5}$) dominates the dynamic behavior and the time delay scales as
\begin{equation}
T_s \sim \dot{\mu}^{-\frac{1}{5}} \Lambda^{\frac{2}{3}} \sim \dot{\alpha}^{-\frac{1}{3}} \nu^{\frac{2}{3}},
\label{timedelayoverdamp}
\end{equation}
and the angle delay is
\begin{equation}
\Delta \alpha = T_s \cdot \dot{\alpha} \sim\dot{\alpha}^{\frac{2}{3}} \nu^{\frac{2}{3}},
\label{angledelayoverdamp}
\end{equation}
Our discrete numerical simulations in later Sections show that these scaling laws of the underlying physics of delayed bifurcation indeed hold.

\section{Static bifurcation in beams and ribbons}
\label{sec:continuousModel}

In this section, we employ AUTO to obtain theoretical predictions of the static bifurcation diagram, and use the dynamic relaxation method together with the discrete model to obtain the numerical solutions of the static snap-through bifurcation of a pre-compressed beam with rotational boundary at one end in 2D space, and a ribbon with transverse shear in 3D space.
The geometric and material parameters of the ribbon considered here are as follows: length $L=1.0$ m, width $W=0.1$ m (which is trivial and can be canceled out), 
thickness $h=0.01$ m, Young's modulus $E=100$ MPa, density $\rho = 1000 \; \mathrm{kg}/\mathrm{m}^3$, and $EI_{1} / EI_{2} = 400.00$, $GJ/EI_{2} = 1.4562$.

\subsection{Snap-through bifurcation in beams}
\label{sec:beamsnapbifurcation}

\begin{figure}[h!]
\centering
\includegraphics[width=1.0\textwidth]{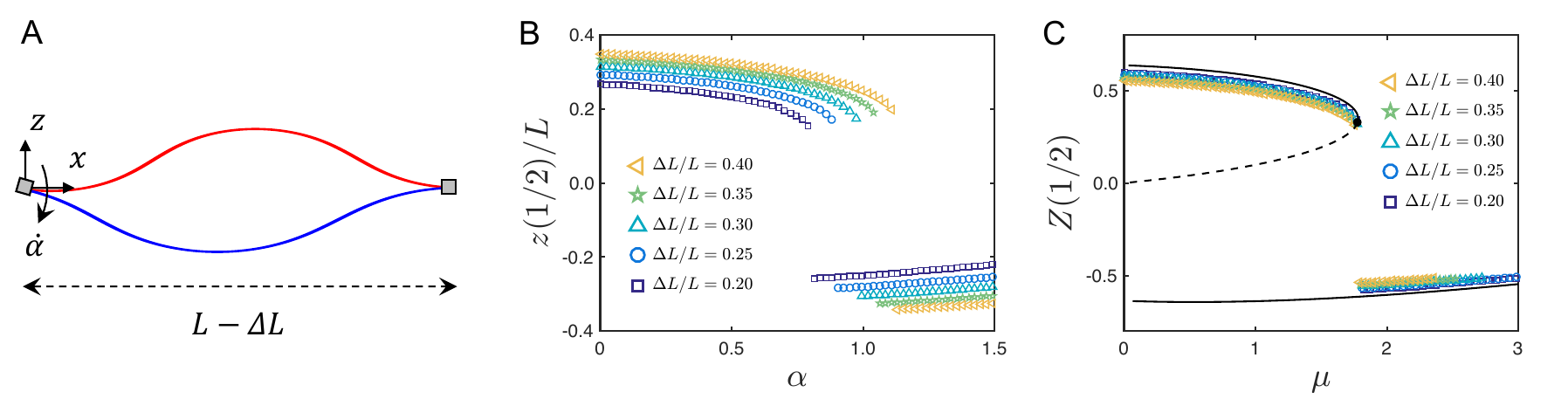}
\caption{Static bifurcation diagrams of an elastic beam. (A) A schematic diagram of the rotation-induced snap-through of beam. (B) The normalized midpoint height, $z(1/2) / L$, as a function of the rotational angle, $\alpha$, for a planner beam with different pre-compression, $\Delta L / L$. (C) The rescaled midpoint height, $Z(1/2) = z(1/2)/ L \cdot (\Delta L / L)^{-1/2} $, as a function of the rescaled rotational angle, $\mu = \alpha \cdot (\Delta L / L)^{-1/2}$. The symbols are from the discrete simulations and the lines from the theoretical solution. The fold points are marked as black dots.}
\label{fig:staticPlanePlot}
\end{figure}

A planar beam experiences buckling instability upon axial compression, and the associated two stable equilibrium configurations would transition from one to the other if one end is rotated by a certain angle (see Fig.~\ref{fig:staticPlanePlot}A), known as snap-through buckling, similar to that in \cite{gomez2017critical} and \cite{liu2021delayed} with different boundary conditions. 
The bifurcation behavior is shown in Fig.~\ref{fig:staticPlanePlot}B, in which the midpoint height of the ribbon normalized by the ribbon length, $z(1/2) / L$, is plotted against the rotational angle, $\alpha$. Five cases with different pre-compressive distances, $\Delta L / L \in \{0.20, 0.25, 0.30, 0.35, 0.40 \}$, are considered. 
The results show that a transition occurs when the midpoint position suddenly jumps from the upper stable branch to the lower stable branch (from the red curve to the blue curve in Fig.~\ref{fig:staticPlanePlot}A) at a critical rotational angle $\alpha_c$. It is clear that the value of the critical rotational angle $\alpha_c$ depends on the pre-compression $\Delta L / L$.

According to the analysis based on the quasi-linear approximation in Sections \ref{sec:Quasilinearapproximation} and \ref{sec:Staticbehavioroflinearproblem}, we re-scale the horizontal and vertical axis as $Z(1/2) = z(1/2) / L \cdot (\Delta L / L)^{-1/2}$ and  $\mu = \alpha \cdot (\Delta L / L)^{-1/2}$, respectively. Fig.~\ref{fig:staticPlanePlot}C shows that results with different pre-compression collapse onto a single master curve, agreeing well with the predictions of the quasi-linear model (Eqs. \eqref{equilibriumconstraint}, \eqref{equilibriumconstraint12})  shown as black curves (the dashed branch is unstable). In addition, we also observe that the jump from upper to lower brunch occurs at a critical value $\mu = \mu_{\fold}$,  corresponding to a saddle–node bifurcation \citep{thompson1990nonlinear,liu2021delayed}. The analytical solution of the bifurcation point has been given in Section \ref{sec:Staticbehavioroflinearproblem} through quasi-linear analysis, which determines that $\mu_{{\fold}}= \pm 1.7818$ and $Z_{{\fold}}(1/2) =  \mp 0.3476$ (see Eq. \eqref{bifurcationsolution}), agreeing with the position of the fold from the nonlinear bifurcation analysis (Fig.~\ref{fig:staticPlanePlot}B). This also indicates that the quasi-linear approach can be employed to study the dynamic problem.
Also, the relative error between the quasi-linear analysis and numerical results decreases for small $\Delta L / L$, because  small deformation is assumed when deriving the theoretical solution in Eq. \eqref{bifurcationsolution}).

\subsection{Snap-through bifurcation in ribbons}
\label{sec:modelValidation}

\begin{figure}[h!]
\centering
\includegraphics[width=0.90\textwidth]{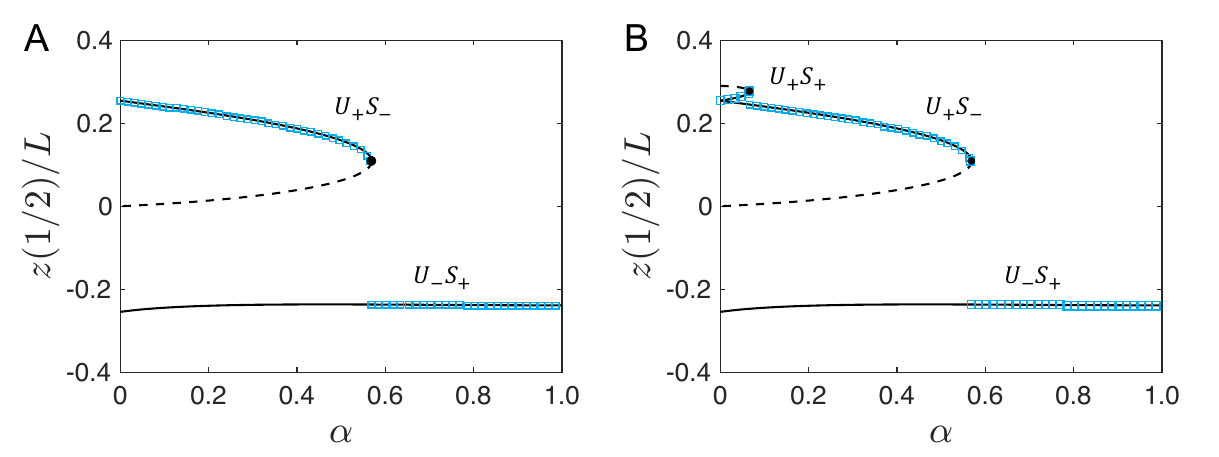}
\caption{Static bifurcation diagrams of an elastic ribbon with pre-compression $\Delta L / L = 0.40$ and pre-shear $\Delta W / L = 0.37$. Snap-through process for the (A) Path 1, $ U_+S_+ \rightarrow U_-S_-$, and, (B) Path 2, $U_+S_- \rightarrow U_+S_+ \rightarrow U_-S_-$. The black solid lines (stable equilibrium) and grey dashed line (unstable equilibrium) are from the theoretical solutions, and the blue symbols are from the discrete numerical simulations. The fold points are marked as black dots.}
\label{fig:autoPlot}
\end{figure}

We further consider the snap-through bifurcation of an elastic ribbon subjected to rotation at one end (see Fig.~\ref{fig:setupPlot}D). The ribbon is first pre-compressed and laterally pre-sheared.
As discussed in Section \ref{sec:problemsetup}, depending on whether the ribbon is in $U_+S_-$ mode or $U_+S_+$ mode upon shearing, there are two snap-through transition paths to reach different static equilibrium configurations.
The ribbon can transition directly from $U_+S_-$ to $U_-S_+$ upon rotation at the left end (Fig.~\ref{fig:setupPlot}E, Fig.~\ref{fig:expSimPlot}A).
Figure~\ref{fig:autoPlot}A shows the bifurcation diagram for this scenario, with results from both the theoretical prediction by AUTO (solid line representing the stable brunch, and the dashed line for the unstable brunch) and the numerical results from the discrete model (symbols).
It is noted that the snap-through is controlled by a single fold point, similar to the planar beam  (see Fig.~\ref{fig:staticPlanePlot}).

If the ribbon is in the $U_+S_+$ mode, it first transitions from $U_+S_+$ back to $U_+S_-$ upon rotation, then jumps to $U_-S_+$ at a larger rotational angle  (Fig.~\ref{fig:setupPlot}F and Fig.~\ref{fig:expSimPlot}B). The bifurcation diagram associated with this snap-through path is shown in Fig.~\ref{fig:autoPlot}B. 
In this case, there are two adjacent fold points, thus the ribbon experiences snap-through transition twice when the end is rotated continuously, with the second one identical to the first scenario (i.e., from $U_+S_-$ to $U_-S_+$). This type of bifurcation is nontrivial and, to our knowledge, has not been reported before.
For both cases, the pre-compression and pre-shear are fixed at $\Delta L / L = 0.4$ and $\Delta W / L = 0.37$, respectively, and the left end is rotated as actuation input. However, the two mirror-symmetric configurations can be manipulated with an input at the right end i.e.,  $U_{+}S_{+}$ will switch directly to $U_{-}S_{-}$, while $U_{+}S_{-}$ will first change to $U_{+}S_{+}$, then eventually to $U_{-}S_{-}$.
%\red{(KJH: do we ever rotate the ribbon at the RIGHT end?)}, \blue{WH: no, we never rotate the right end.}

\begin{figure}[ht!]
\centering
\includegraphics[width=0.90\textwidth]{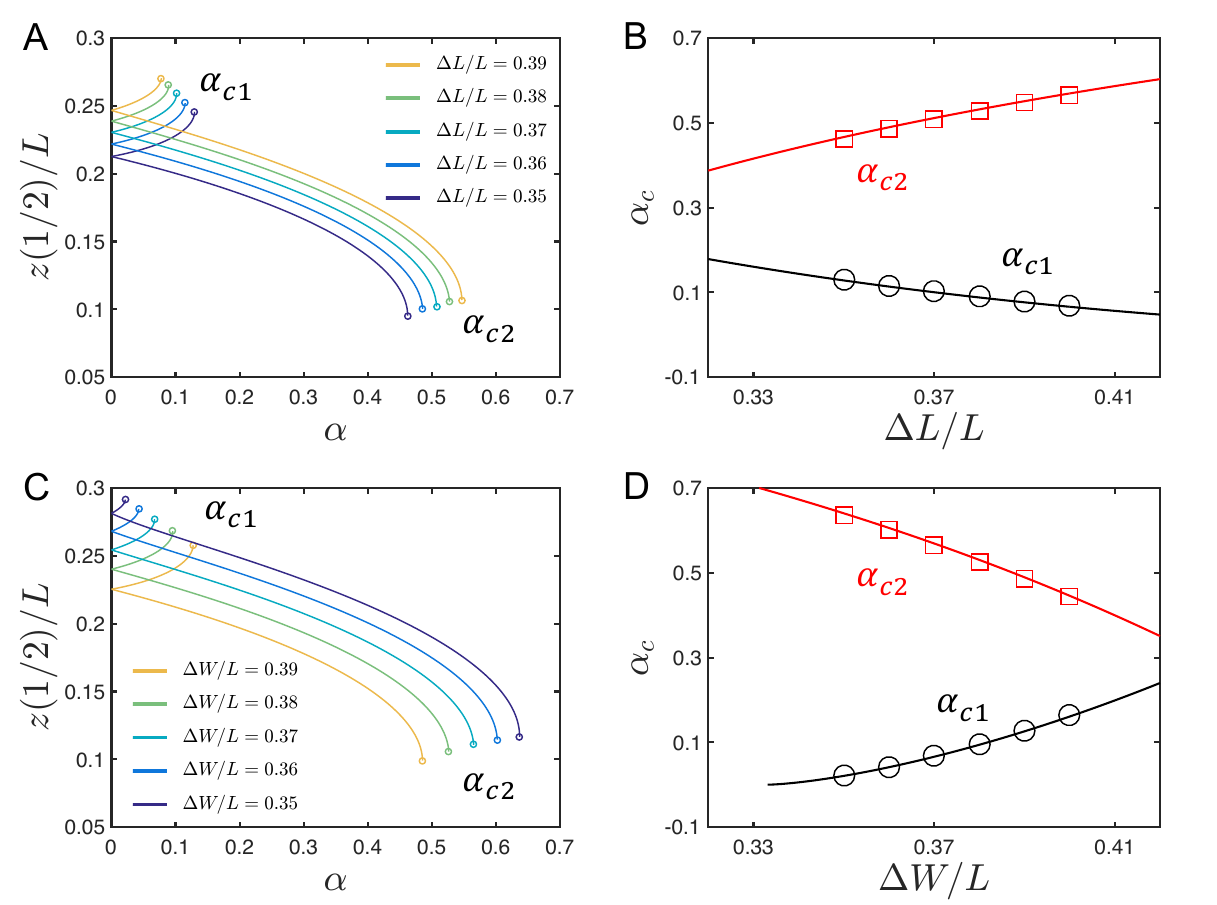}
\caption{Parametric analysis for the static bifurcation diagrams of an elastic ribbon. (A) The normalized midpoint height, $z(1/2) / L$, as a function of the rotational angle, $\alpha$, for the ribbon with fixed pre-shear, $\Delta W / L = 0.37$, and different pre-compression, $\Delta L / L \in [0.35, 0.39]$. (B) The critical rotational angle, $\alpha_{c}$, as a function of the pre-compression, $\Delta L / L$. (C) The normalized midpoint height, $z(1/2) / L$, as a function of the rotational angle, $\alpha$, for the ribbon with fixed pre-compression, $\Delta L / L = 0.40$, and different pre-shear, $\Delta W / L \in [0.35, 0.39]$. (D) The critical rotational angle, $\alpha_{c}$, as a function of the pre-shear, $\Delta W / L$. In (B) and (D), the solid lines are from the theoretical model, and the symbols are obtained from the discrete simulation.}
\label{fig:compressShearPhasePlot}
\end{figure}

Similar to the beam snapping case, the fold point for ribbons depends on the amount of pre-compression and pre-shear. A parametric study is carried out to understand the dependence of the two fold points (corresponding to the case with two snap-through transitions) on the compression and shear (see Fig.~\ref{fig:compressShearPhasePlot}).
For fixed shear $\Delta W / L = 0.37$ and the pre-compression $\Delta L / L$ ranging from $0.35$ to $0.39$, the relationship between the normalized midpoint height, $z(1/2)/L$, and the rotational angle, $\alpha$, is plotted in Fig.~\ref{fig:compressShearPhasePlot}A.
Note that there exist two fold points, resulting in two critical angles $\alpha_{c1}$ and $\alpha_{c2}$. In Fig.~\ref{fig:compressShearPhasePlot}B, the critical angles, $\alpha_{c1}$ and $\alpha_{c2}$, are plotted against the pre-compression, $\Delta L / L$, from the results using the discrete numerical simulations (symbols) and the theoretical analysis (solid lines). They agree with each other well.
For fixed pre-compression $\Delta L / L = 0.40$ and a pre-shear ranging from $0.35$ to $0.39$, the relationship between the normalized midpoint height, $z(1/2)/L$, and the rotational angle, $\alpha$, is shown in Fig.~\ref{fig:compressShearPhasePlot}C, and the dependence of the critical rotational angles, $\alpha_{c1}$ and $\alpha_{c2}$, on the pre-shear, $\Delta W / L$, are given in Fig.~\ref{fig:compressShearPhasePlot}D. Again, the theoretical predictions (solid lines) and the discrete simulation results (symbols) agree well.

\section{Dynamic snap-through}
\label{sec:dynamicsnapthrough}

In this section, we move to consider the dynamics of the snap-through instabilities, for both beams in 2D space and ribbons in 3D space, where both inertia and damping effects are accounted for.

\subsection{Dynamic snap-through of beams}

The dynamic snap-through of a planar beam in 2D space is studied using the theoretical model and discrete numerical simulations.
A beam with pre-compression $\Delta L / L =0.40$ is considered. The snap-through behavior triggered by rotating one end with constant angular speed $\dot{\alpha}$, i.e., $\phi(0,t) = \alpha (t) = \alpha_0 + \dot{\alpha}t$ is investigated, where $\alpha_0 = 0$ and $\dot{\alpha}$ correspond to the initial angle and the angular velocity of the rotating end, respectively. %
The dynamic behavior of the snap-through instability of the beam is simulated with different damping coefficient, $\nu$, and different rotating rate, $\dot{\alpha}$.

\begin{figure}[h!]
\centering
\includegraphics[width=0.90\textwidth]{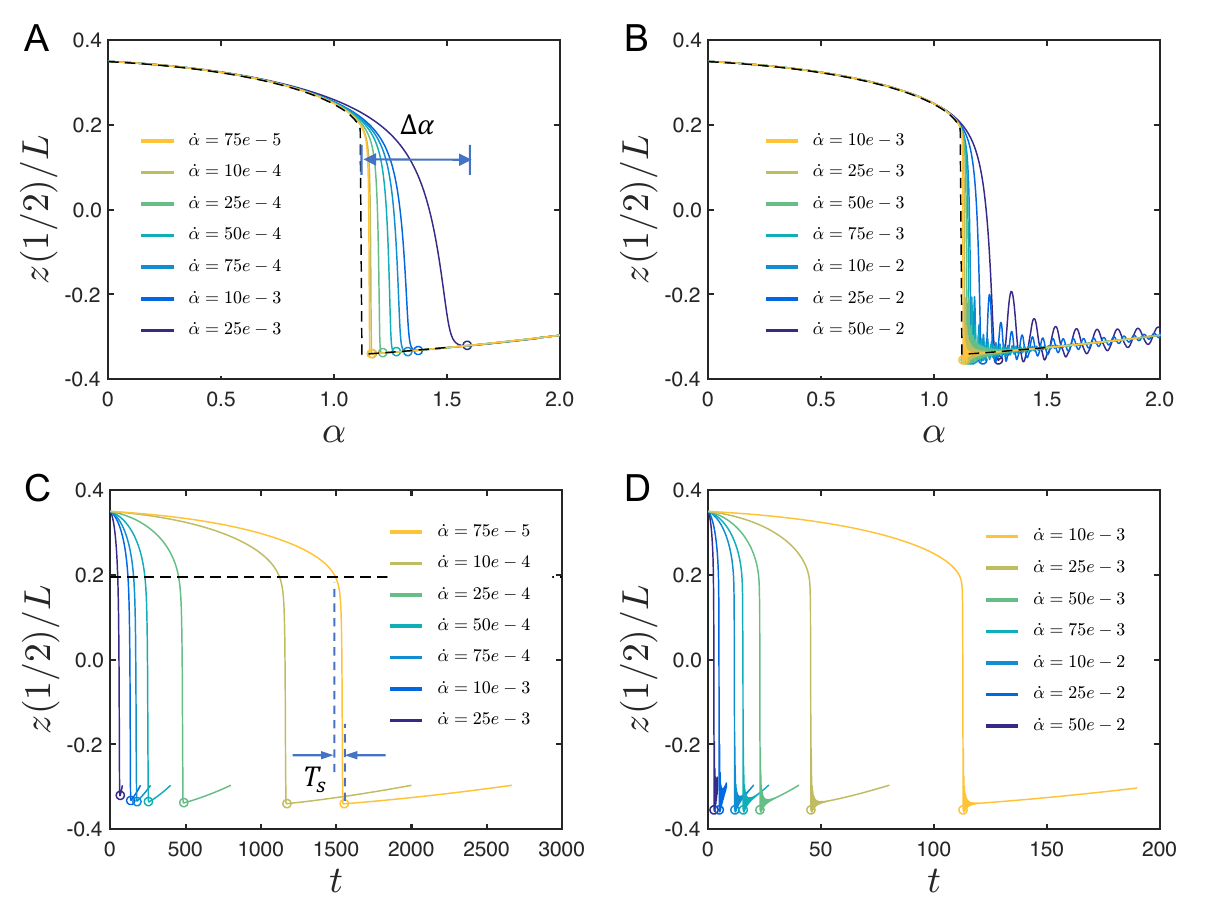}
\caption{Dynamic snap through of a planar beam with fixed pre-compression, $\Delta L / L = 0.4$. The normalized midpoint height, $z(1/2) / L$, as a function of the rotational angle, $\alpha$, for: (A) Overdamping case, $\nu = 1e3$ and, (B) underdamping case, $\nu = 1e-3$. The black dashed line is from static analysis. The normalized midpoint height, $z(1/2) / L$, as a function of the time, $t$, for (C) Overdamping case ($\nu = 1e3$) and, (D) underdamping case ($\nu = 1e-3$). The angle delay, $\Delta \alpha$, and the time delay, $T_s$, are marked in (A) and (C), respectively.}
\label{fig:planeDynamicPlot}
\end{figure}

Figures ~\ref{fig:planeDynamicPlot}A and B show the normalized midpoint height of the ribbon, $z(1/2) / L$, as a function of the rotational angle, $\alpha = \dot{\alpha} t$, with different rotating rate $\dot{\alpha}$, for both the overdamping case (with damping coefficient $\nu = 1e3$) and the underdamping case (with $\nu = 1e-3$), respectively.
The results show that there is significant delay in the dynamic snap-through instability, i.e., the snap-through occurs at delayed positions (the value of $\alpha$) away from the fold point predicted by the quasi-static theory. In Fig. ~\ref{fig:planeDynamicPlot}A, the difference between the actual snap-through point and the quasi-static fold point is defined as the angle delay, $\Delta \alpha$. The magnitude of this delay $\Delta \alpha$ decreases as $\dot{\alpha} \rightarrow 0$ for both overdamped and underdamped cases.
We further note that, for the overdamped case in Fig.~\ref{fig:planeDynamicPlot}A, the transition between two stable equilibrium configurations is smooth; while for the underdamped case, significant oscillations can be observed after the snap-through transition, as shown in Fig.~\ref{fig:planeDynamicPlot}B.
In Figs.~\ref{fig:planeDynamicPlot}C and D, we plot the relationship between the normalized midpoint height, $z(1/2) / L $, and the loading time, $t$.
The time delay during the snap-through process (marked in Fig.~\ref{fig:planeDynamicPlot}C) can be calculated by $T_{s} =  {\Delta \alpha} / {\dot{\alpha}} $, and it increases as $\dot{\alpha} \rightarrow 0$.

\begin{figure}[h!]
\centering
\includegraphics[width=1.0\textwidth]{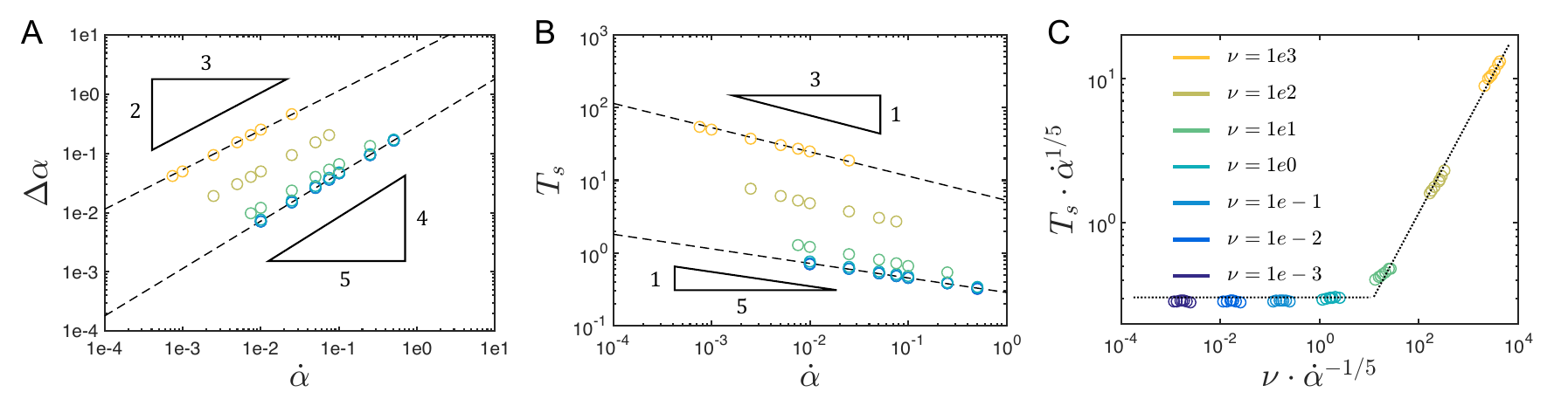}
\caption{Scaling of the delayed bifurcation in dynamic snap-through of an elastic beam. (A) Scaling law between the angle delay, $\Delta \alpha$, and the loading rate, $\dot{\alpha}$, for different damping coefficient, $\nu$. (B) Scaling law between the time delay, $T_{s}$, and the loading rate, $\dot{\alpha}$, for different damping coefficient, $\nu$. (C) Duration of the snap-through, $\mathcal{T}_s \sim T_s \cdot \dot{\alpha}^{1/5}$, as a function of the damping ratio, $\Lambda \sim \nu \cdot \dot{\alpha}^{-1/5}$.}
\label{fig:planeScalePlot}
\end{figure}

The delay of the dynamic snap-through $\Delta \alpha$ is plotted as a function of the rotating rate $\dot{\alpha}$ in Fig.~\ref{fig:planeScalePlot}A for seven  different values of damping coefficient, $\nu \in [1e-3, 1e3]$. The results all fall onto straight lines in a log-log plot, indicating a power-law dependence 
%\red{(Note: a linear dependence in log-log plot is power-law dependence. Straight line in a log-linear plot is exponential.)}
between $\Delta \alpha$ and $\dot{\alpha}$. The asymptotic analysis in Section \ref{sec:Snapthroughdynamicsoflinearproblem} indicates that the power exponents for the $\Delta \alpha$ -- $\dot{\alpha}$ relations for underdamped and overdamped limits should be ${4}/{5}$ (Eq. \eqref{timedelayunderdamp}) and ${2}/{3}$  (Eq. \eqref{angledelayunderdamp}), respectively, consistent with our numerical simulation results in Fig.~\ref{fig:planeScalePlot}A.  The asymptotic analysis results in Eqs. \eqref{timedelayoverdamp} and \eqref{angledelayoverdamp} dictates that the scalings for time delay $T_s$ with respect to $\dot{\alpha}$ are $-{1}/{5}$ and $-{1}/{3}$ for over- and underdamped cases, respectively, as demonstrated in Fig.~\ref{fig:planeScalePlot}B. Furthermore, the asymptotic analysis of Sections \ref{sec:Snapthroughdynamicsoflinearproblem} suggests that these  numerical data should collapse when  rescaled by $\mathcal{T}_s \sim T_s \cdot \dot{\alpha}^{1/5}$ and plotted as a function of the normalized damping coefficient $\Lambda \sim \nu \cdot \dot{\alpha}^{-1/5}$. Figure~\ref{fig:planeScalePlot}C indeed shows the collapsed data points. The two asymptotic regimes, where $\Lambda \ll 1$ and $\Lambda \gg 1$, are clearly demarcated by the dashed lines in the figure.

\subsection{Dynamic snap-through of ribbons}

\begin{figure}[h!]
\centering
\includegraphics[width=0.90\textwidth]{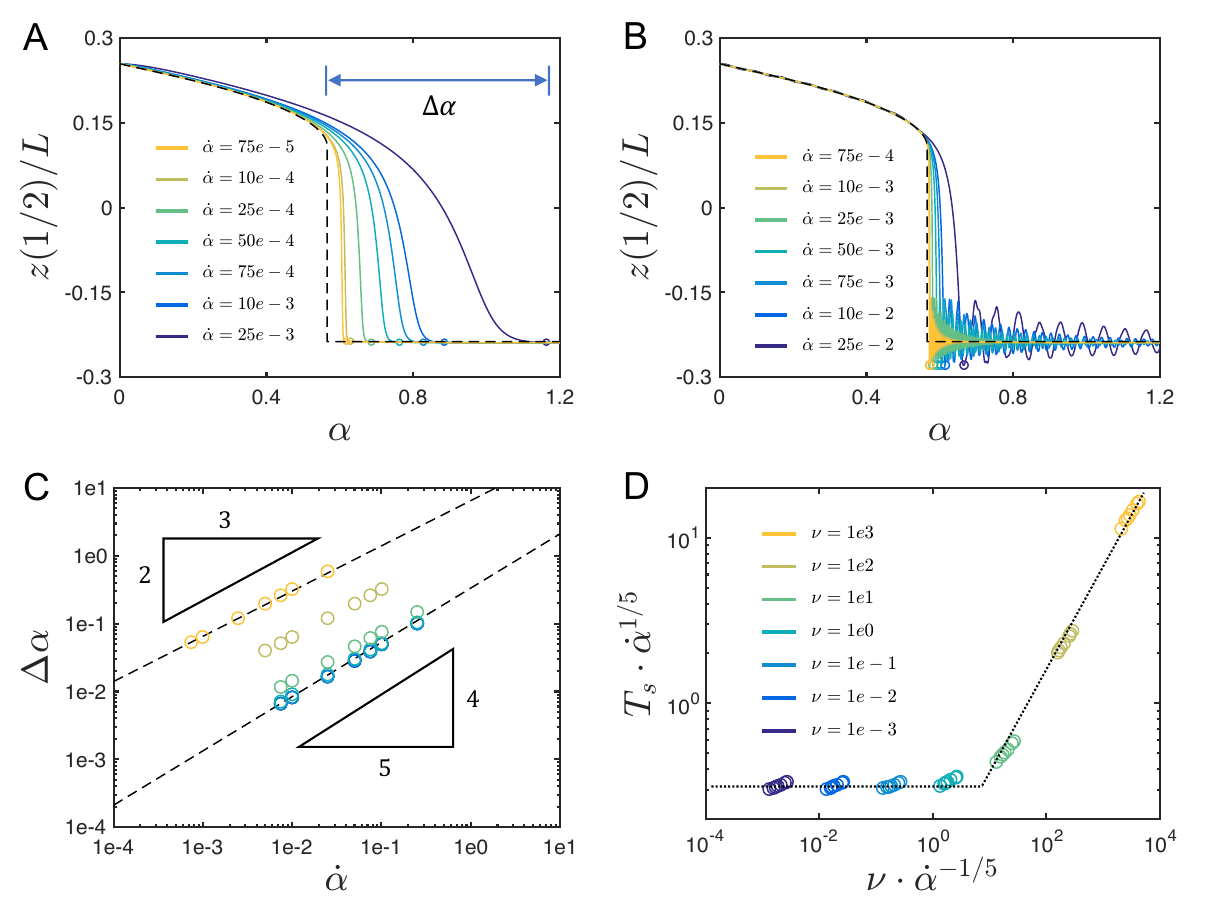}
\caption{Dynamic snap-through of an elastic ribbon with pre-compression, $\Delta L / L = 0.40$, and pre-shear, $\Delta W / L = 0.37$, through the Path $U_+S_- \rightarrow U_-S_+$ for: (A) Overdamping case, $\nu = 1e3$ and, (B) Underdamping case, $\nu = 1e-3$. The black dashed line is from the static analysis. (C) Scaling law between the angle delay, $\Delta \alpha$, and the loading rate, $\dot{\alpha}$, for different damping coefficient, $\nu$. (D) Duration of the snap-through, $\mathcal{T}_s \sim T_s \cdot \dot{\alpha}^{1/5}$, as a function of the damping ratio, $\Lambda \sim \nu \cdot \dot{\alpha}^{-1/5}$.}
\label{fig:case1DynamicPlot}
\end{figure}

For dynamic snap-through of ribbons, we focus on the case of pre-compression $\Delta L / L =0.40$ and pre-shear $\Delta L / L =0.37$.
We first consider the snap-through path, $ U_+S_- \rightarrow U_-S_+$ using the discrete numerical method, and record the evolution of ribbon configuration for different rotational rate $\dot{\alpha}$.
Figs~\ref{fig:case1DynamicPlot}A and B show the dependence of the normalized midpoint height, $z(1/2) / L$, on the rotational angle, $\alpha$, for both overdamped ($\nu = 1e3$) and underdamped ($\nu = 1e-3$) cases, respectively. Similar to the snap-through of beams, a significant delay is found in the ribbon, and the smooth and oscillatory behaviors are present in the overdamped and underdamped  cases. The angle delay $\Delta \alpha$ for the ribbons with different damping coefficient $\nu$ as a function of the rotation rate $\dot{\alpha}$ is shown in Fig.~\ref{fig:case1DynamicPlot}C. The linear relationship of all data points in the log-log plot indicates a power-law scaling between $\Delta \alpha$ and $\dot{\alpha}$ with the power exponents of ${4}/{5}$ and ${2}/{3}$ for underdamped and overdamped limits, respectively. These results indicate that the introduction of pre-shear (from beam to ribbon) does not affect the intrinsic feature of the delay in dynamic snap-through of elastic slender objects, be it a slender beam or a thin ribbon. Accordingly, the time delay $T_s$ can be calculated and rescaled as $\mathcal{T}_s \sim T_s \cdot \dot{\alpha}^{1/5}$, and plotted  as a function of damping ratio $\Lambda$ as in Fig.~\ref{fig:case1DynamicPlot}D, again leading to a collapsed master curve with two  asymptotic regimes ($\Lambda \ll 1$ and $\Lambda \gg 1$), similar to that of a beam snap-through in Fig.~\ref{fig:planeScalePlot}C.

\begin{figure}[t!]
\centering
\includegraphics[width=0.90\textwidth]{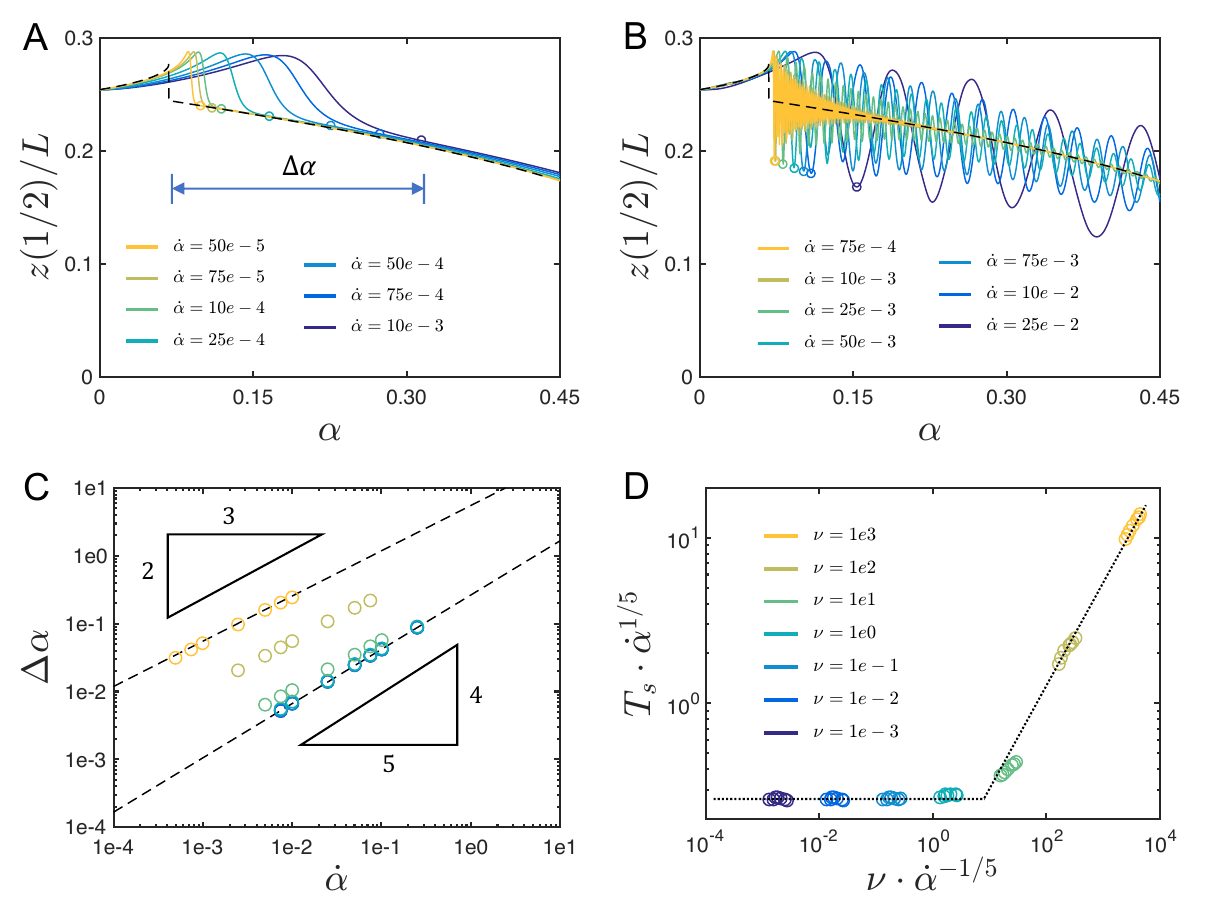}
\caption{Dynamic snap through of an elastic ribbon with pre-compression, $\Delta L / L = 0.40$, and pre-shear, $\Delta W / L = 0.37$, through the Path $U_+S_+ \rightarrow U_+S_-$ for: (A) Overdamping case, $\nu = 1e3$ and, (B) Underdamping case, $\nu = 1e-3$. The black dashed line is from the static analysis. (C) Scaling law between the angle delay, $\Delta \alpha$ and the loading rate, $\dot{\alpha}$, for different damping coefficientm $\nu$. (D) Duration of the snap-through, $\mathcal{T}_s \sim T_s \cdot \dot{\alpha}^{1/5}$, as a function of the damping ratio, $\Lambda \sim \nu \cdot \dot{\alpha}^{-1/5}$.}
\label{fig:case2DynamicPlot}
\end{figure}

We then consider the dynamic snap-through of elastic ribbon of the nontrivial path, i.e., $U_+S_+ \rightarrow U_+S_- \rightarrow U_-S_+$. Since the second snap-through transition in this path ($U_+S_- \rightarrow U_-S_+$) is identical to the path considered earlier, here we focus on the first transition, i.e., $U_+S_+ \rightarrow U_+S_-$.
Figures~\ref{fig:case2DynamicPlot}A and B show the normalized midpoint height of the ribbon, $z(1/2)/L$, versus the rotational angle, $\alpha$, for both overdamped ($\nu = 1e3$) and underdamped ($\nu = 1e-3$) situations. This transition also shows significantly delayed under dynamic loading. Figure~\ref{fig:case2DynamicPlot}C shows the dependence of the angle delay, $\Delta \alpha$, on the loading rate, $\dot{\alpha}$, for different damping coefficient, $\mu \in \{1e3, 1e-3\}$. The results show clear power-law scaling, with the power exponents of ${4}/{5}$ and ${2}/{3}$, respectively. Again, the rescaled data in Fig.~\ref{fig:case2DynamicPlot} D, not surprisingly, show collapse onto a master curve with asymptotic trends.
These results show that all snap-through transitions in both Path 1 and Path 2 are governed by the fold point bifurcation and share the same intrinsic scaling laws.

\section{Mode skipping and selection enabled by delayed bifurcation}
\label{sec:PathControl}

The Path 2 transition discussed above ($U_+S_+ \rightarrow U_+S_- \rightarrow U_-S_+$), due to the existence of multiple fold points, provides the possibility to manipulate the snap-through behavior of ribbons, e.g., to achieve skipping or selection of particular transition modes, by controlling the dynamic loading conditions.

\subsection{Mode skipping}

\begin{figure}[t!]
\centering
\includegraphics[width=0.90\textwidth]{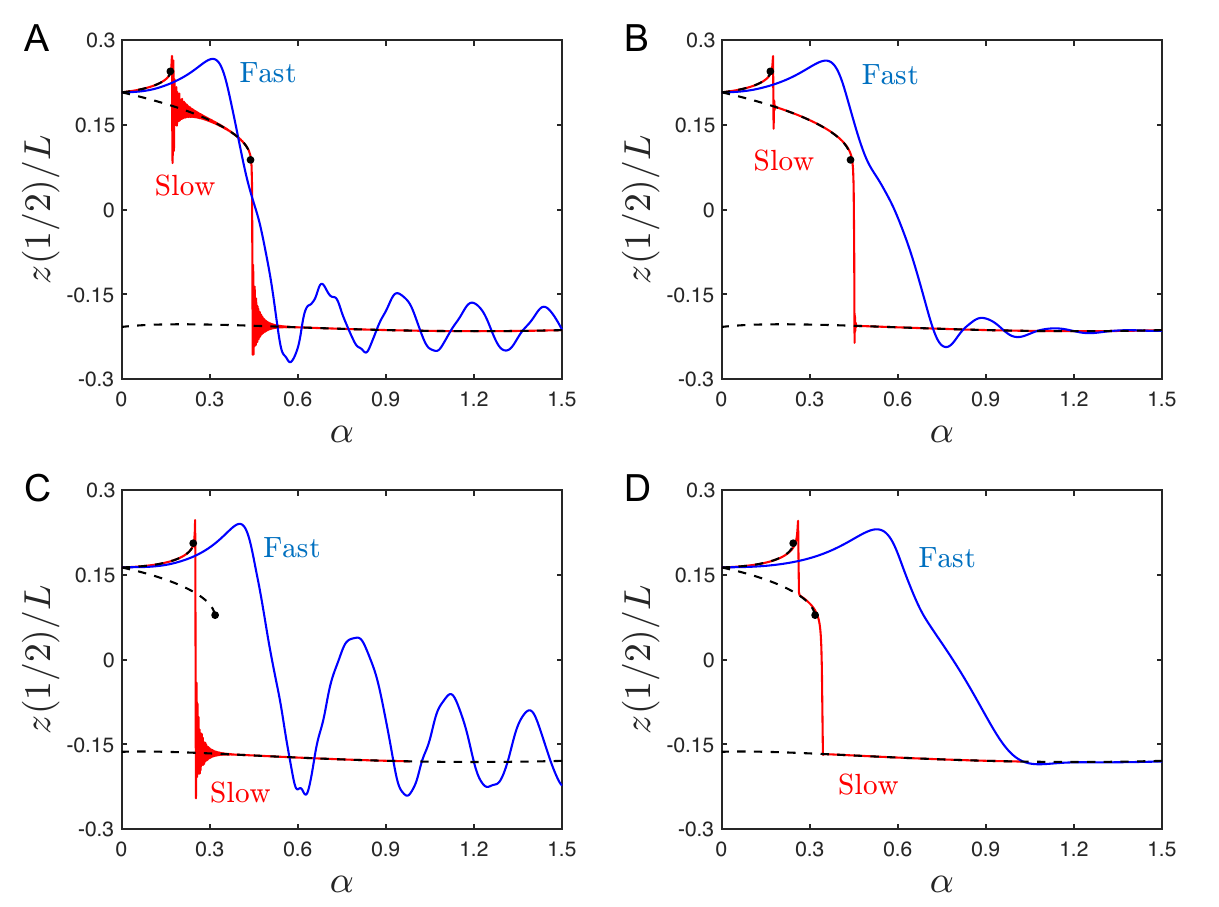}
\caption{Mode skipping in dynamic snap-through of  elastic ribbons with pre-shear $\Delta W / L = 0.40$. The normalized midpoint height, $z(1/2) / L$, as a function of the rotational angle, $\alpha$, and different damping coefficients of (A) $\nu=0.1$ and (B) $\nu=10.0$ (with pre-compression $\Delta L / L = 0.40$); and (C) $\nu=0.3$ and (D) $\nu=30.0$ (with $\Delta L / L = 0.35$). The fast and slow cases correspond to the rotational rate $\dot{\alpha} = 1.00$ and $0.01$ $\mathrm{rad}/\mathrm{s}$, respectively, for all cases. The black dashed line is from the static analysis.
}
\label{fig:pathPlot}
\end{figure}

\begin{figure}[h!]
\centering
\includegraphics[width=1.00\textwidth]{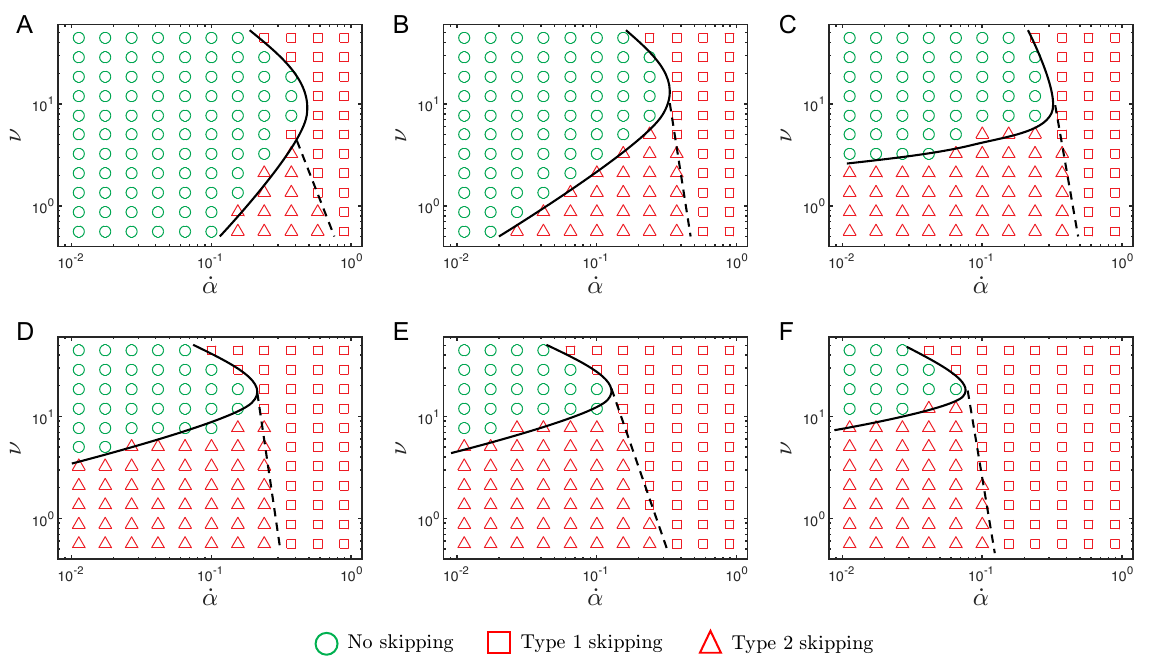}
\caption{Phase diagram of mode skipping in the 2D space of ($\dot{\alpha}, \nu$) for elastic ribbons with fixed pre-shear $\Delta W / L = 0.40$, and the pre-compression is set as: (A) $\Delta L /L = 0.40$, (B) $\Delta L /L = 0.39$, (C) $\Delta L /L = 0.38$, (D) $\Delta L /L = 0.37$, (E) $\Delta L /L = 0.36$, (F) $\Delta L /L = 0.35$. The phases of no skipping (green circles) and with skipping (red symbols) are separated by the solid line; and the sub-phases of `type 1 skipping' (squares) and `type 2 skipping' (triangles) are separated by dashed lines.}
\label{fig:pathPhasePlot}
\end{figure}

Figure~\ref{fig:pathPlot}A shows the normalized midpoint height of the elastic ribbon, $z(1/2)/ L$, as a function of the rotational angle, $\alpha$, for a ribbon (with pre-compression $\Delta L / L = 0.40$, pre-shear $\Delta W / L= 0.40$, and low damping coefficient $\nu = 0.1$) subjected to either low rotational rate of $\dot{\alpha} = 0.01$ rad/s, or high rotational rate of $\dot{\alpha} = 1.00$ rad/s.
As expected, for low rotational rate (red line), there are two snap-through jumps during the loading process (from $U_+S_+$ to $U_+S_-$, and then to $U_-S_+$), agreeing with the prediction from the quasi-static analysis (dashed black lines). At high rotational rate, however, the first snap-through jump ($U_+S_+ \rightarrow U_+S_-$) is skipped due to the delay under dynamic loading (blue line).
Under high damping ($\nu = 10.0$), Fig.~\ref{fig:pathPlot}B shows a similar behavior, while the oscillations after the jump are significantly suppressed.
This mode skipping phenomenon is the consequence of sufficiently large delay in snap-through induced by high dynamic loading to jump over the gap between the first and second transitions.

Furthermore, by reducing the pre-compression to $\Delta L / L = 0.35$ while keeping the pre-shear as $\Delta W / L= 0.40$, we find a new type of mode skipping. For low damping ($\nu = 0.3$),  Fig.~\ref{fig:pathPlot}C shows that, under fast loading ($\dot{\alpha} = 0.01$ rad/s, blue curve), the ribbon still jumps from $U_+S_+$ to $U_-S_+$ directly. But the slow loading case behaves differently. It goes beyond $U_+S_+$, passes through the $U_+S_-$ mode, and jumps to $U_-S_+$ instead of stopping at $U_+S_-$. This is because the elastic energy release during transition from $U_+S_+$ to $U_+S_-$ is sufficient to overcome the energy barrier between $U_+S_-$ and $U_-S_+$. At high damping $\nu = 30.0$, Fig.~\ref{fig:pathPlot}D shows that part of the released elastic energy is dissipated by the damping effect, and the $U_+S_-$ cannot be skipped (red line). Under fast loading, e.g., $\dot{\alpha} = 1.00$ rad/s (the blue line), a similar mode skipping phenomenon caused by the delay bifurcation exists.

The above analyses can be compiled into a comprehensive phase diagram of mode skipping behavior for a ribbon with different geometries and damping coefficients under different loading rates, as shown in Fig.~\ref{fig:pathPhasePlot}.
Here, the pre-shear is fixed as $\Delta W / L = 0.40$, and the pre-compression varies as $\Delta L / L \in \{0.40, 0.39, 0.38, 0.37, 0.36, 0.35 \}$ in Fig.~\ref{fig:pathPhasePlot}A to Fig.~\ref{fig:pathPhasePlot}F. The detailed analysis for the two extreme cases (i.e., $\Delta L / L = 0.40$ and $\Delta L / L = 0.35$) are already presented in Fig.~\ref{fig:pathPlot}.
Figure~\ref{fig:pathPhasePlot} generally indicates, for given $\Delta L / L$ and $\Delta W / L$, in the $\dot{\alpha} - \nu$ space, a ribbon exhibits either the skipping behavior (red symbols) or no-skipping behavior (green symbols) with the black solid line representing the phase boundary. For the skipping phase, two sub-phases indicating different types of skipping mechanisms can be identified, i.e., the direct jump caused by the energy release referred to as Type $1$ skipping (e.g., red curves in Fig.~\ref{fig:pathPlot}C), or the jump induced by the delay bifurcation referred to as Type $2$  (e.g., blue curves in Fig.~\ref{fig:pathPlot}A). To determine the skipping type, we evaluate the location the first local minima in the $z(1/2)/L \sim \alpha$ plot (Fig.~\ref{fig:pathPlot}): if it is smaller than the second critical snap angle, $\alpha_{c2}$, we refer it as `Type $1$ skipping' (red squares), otherwise, it is the `Type $2$ skipping' (red triangles). These sub-regions are  separated by the dashed line.

\subsection{Mode selection}
\label{sec:modelSelection}

The static bifurcation diagram of the snap-through Path 2 ($U_+S_+ \rightarrow U_+S_- \rightarrow U_-S_+$, Fig.~\ref{fig:autoPlot}) indicates that both $U_{+}S_{-}$ and $U_{-}S_{+}$ are stable equilibrium configurations when the rational angle $\alpha$ is  between of the first and the second fold points, $\alpha_{c1} < \alpha < \alpha_{c2}$.
This feature allows the possibility of mode selection by controlling the dynamic loading.

\begin{figure}[t!]
\centering
\includegraphics[width=0.90\textwidth]{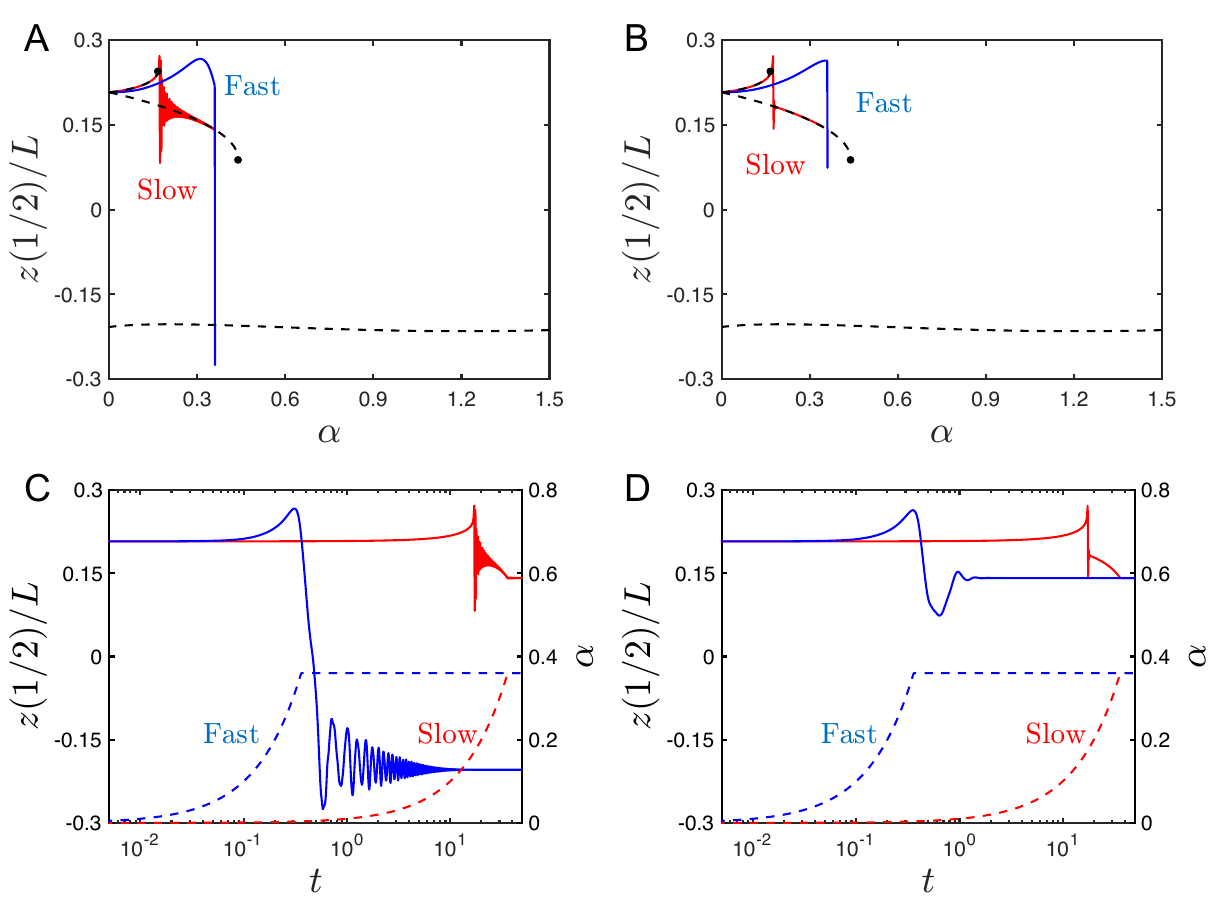}
\caption{Mode selection in dynamic snap-through of an elastic ribbon with pre-compression $\Delta L / L = 0.40$ and pre-shear $\Delta W / L = 0.40$. The normalized midpoint height, $z(1/2) / L$, as a function of the rotational angle, $\alpha$, for different damping coefficients: (A) $\nu=0.1$ and (B) $\nu=10.0$ (the black dashed line is from static analysis). The normalized midpoint height, $z(1/2) / L$, and the rotational angle, $\alpha$, as a function of the time, $t$, for different damping coefficients: (C) $\nu=0.1$ and (D) $\nu=10.0$. The fast and slow cases correspond to the rotational rate $\dot{\alpha} = 1.00$ and $0.01$ $\mathrm{rad}/\mathrm{s}$, respectively, for all cases. The stopping position of the rotational angle is $\alpha_{\mathrm{stop}} = 0.36$.}
\label{fig:modePlot}
\end{figure}

For a ribbon with fixed pre-compression $\Delta L / L = 0.40$ and pre-shear $\Delta W / L = 0.40$, we can dynamically rotate one end and stop at $\alpha_{\mathrm{stop}} = 0.36$, between of $\alpha_{c1}$ and $\alpha_{c2}$. The results are shown in Fig.~\ref{fig:modePlot}.
At low damping  ($\nu = 0.1$), Figs.~\ref{fig:modePlot}A shows that, for small rotational rate, $\dot{\alpha} = 0.01$ (the red line), the ribbon stays at the $U_{+}S_{-}$ mode (corresponding to $z(1/2) > 0$) due to the energy barrier between the two stable configurations.
For high rotational rate $\dot{\alpha} = 100$ (blue line), however, the kinetic energy stored in the ribbon is sufficient to overcome the energy barrier to result in a jump to the other stable equilibrium, $U_{-}S_{+}$ ($z(1/2) < 0$), even though the rotational angle is smaller than that of the second fold point.
At high damping  ($\nu = 10.0$), Fig.~\ref{fig:modePlot}B shows the kinetic energy is dissipated and cannot help the system overcome the energy barrier. As a result, both low and high loading rates lead to the upper equilibrium mode ($U_{+}S_{-}$).
The physical process of the above cases are further clarified in Fig.~\ref{fig:modePlot}C and D, in which the rotational angle, $\alpha$ (representing the loading process), and the midpoint height, $z(1/2) / L$ (representing the response), are plotted versus time, $t$, showing that the ribbon may reach either different or the same configurations at the same final rotational angle with different loading rates. 

\begin{figure}[t!]
\centering
\includegraphics[width=0.90\textwidth]{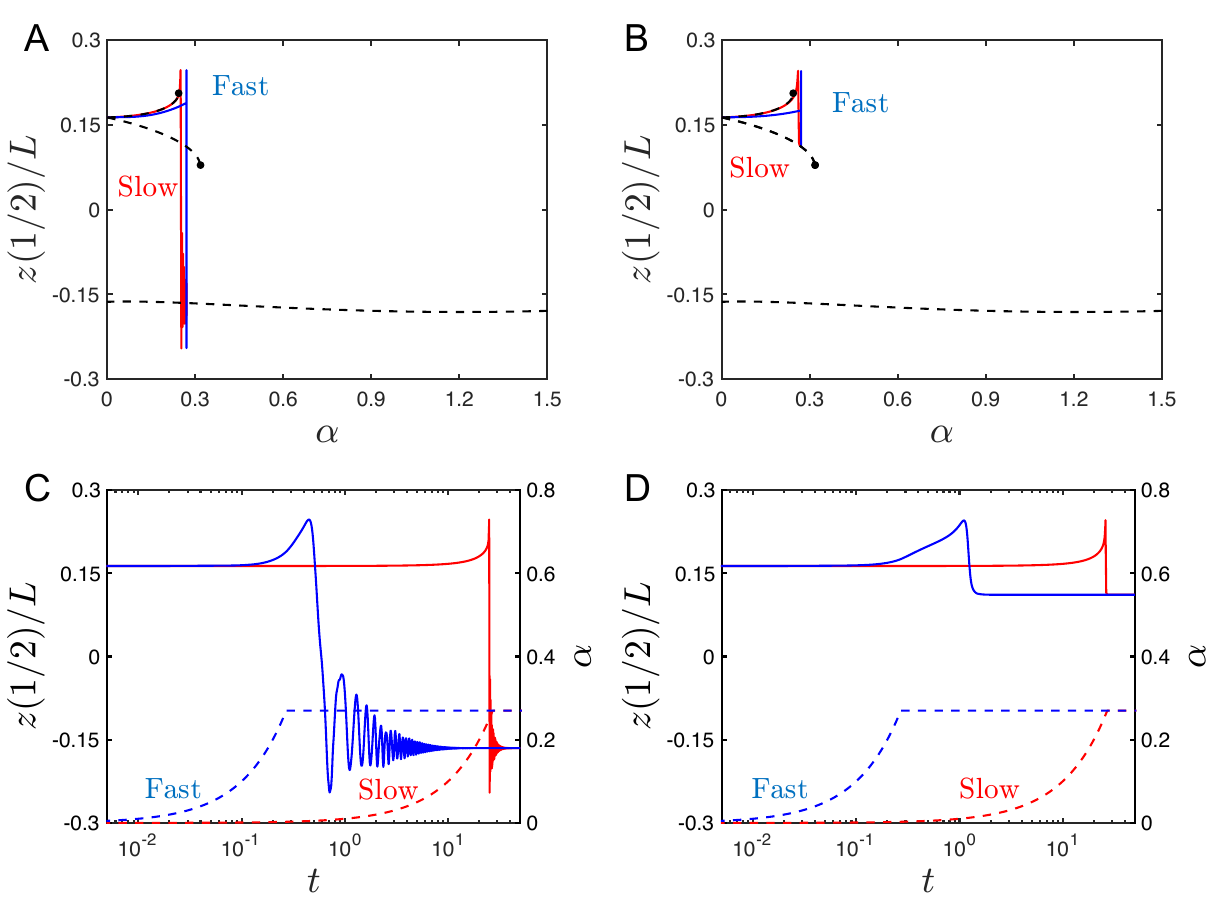}
\caption{Mode selection in dynamic snap-through of an elastic ribbon with pre-compression $\Delta L / L = 0.35$ and pre-shear $\Delta W / L = 0.40$. The normalized midpoint height, $z(1/2) / L$, as a function of the rotational angle, $\alpha$, for different damping coefficients: (A) $\nu=0.3$ and (B) $\nu=30.0$ (the black dashed line is from static analysis). The normalized midpoint height, $z(1/2) / L$, and the rotational angle, $\alpha$, as a function of the time, $t$, for different damping coefficients: (C) $\nu=0.3$ and (D) $\nu=30.0$. The fast and slow cases correspond to the rotational rate $\dot{\alpha} = 1.00$ and $0.01$ $\mathrm{rad}/\mathrm{s}$, respectively, for all cases. The stopping position of the rotational angle is $\alpha_{\mathrm{stop}} = 0.27$.}
\label{fig:mode2Plot}
\end{figure}

However, with a different ribbon pre-compression $\Delta L / L = 0.35$, the same pre-shear $ \Delta W / L = 0.40$, and the final rotational angle  $\alpha_{\mathrm{stop}} = 0.27$ ($\alpha_{c1} < \alpha_{\mathrm{stop}} < \alpha_{c2}$),
the $U_{+}S_{+}$ mode jumps directly to $U_{-}S_{+}$ at low damping ($\nu=0.3$) for both high and low loading rates ($\dot{\alpha} \in \{0.01, 1 \}$ rad/s), because the energy released during the first snap-through ($U_{+}S_{+} \rightarrow U_{+}S_{-}$) can overcome the energy barrier between the $U_{+}S_{-}$ and $U_{-}S_{+}$, as shown in Fig.~\ref{fig:mode2Plot}A and C.
On the other hand, the ribbons may both remain in the $U_{+}S_{-}$ mode if the system has high damping (e.g., with $\nu=30$), as the energy during the first snap would dissipate and fail to overcome the barrier between the $U_{+}S_{-}$ and $U_{-}S_{+}$, as shown in Figs.~\ref{fig:mode2Plot}B and D.

\begin{figure}[t!]
\centering
\includegraphics[width=0.9\textwidth]{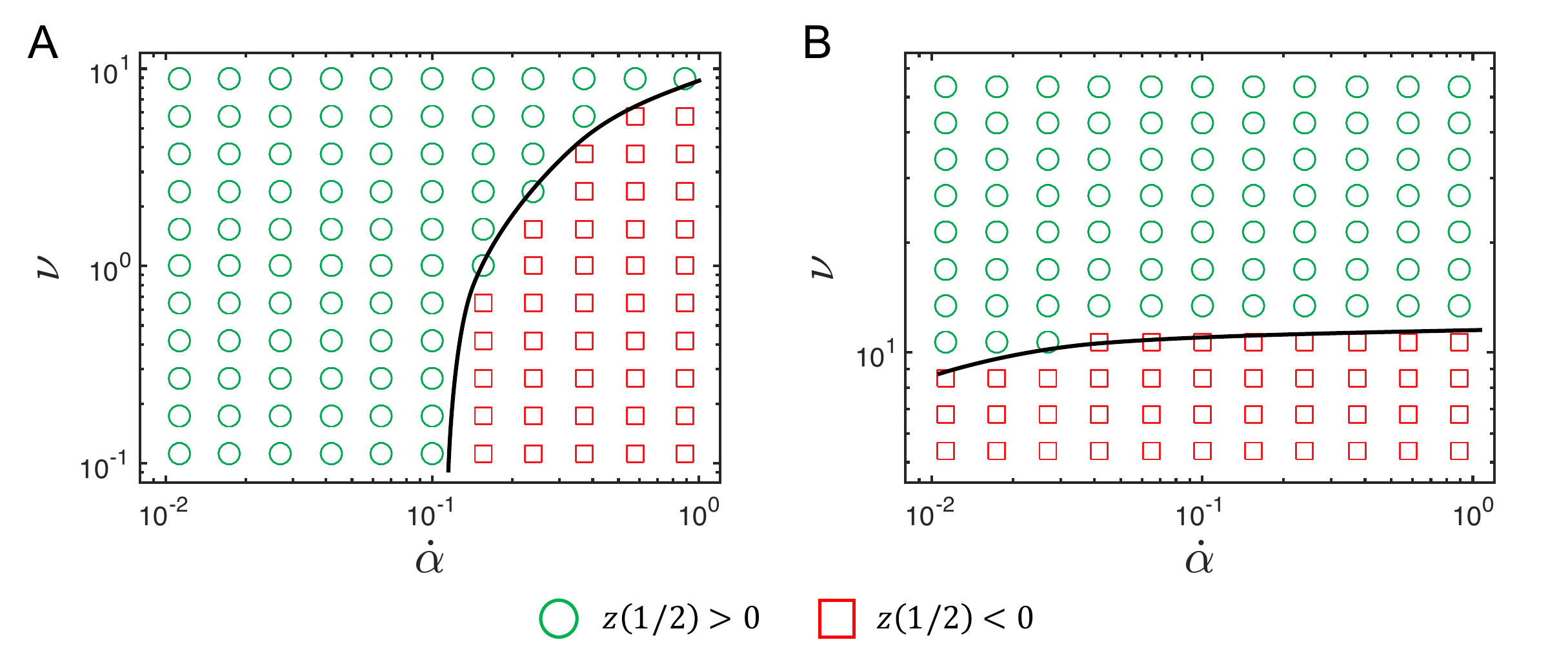}
\caption{Phase diagram of mode selection in the 2D space of ($\dot{\alpha}, \nu$) for elastic ribbon with fixed pre-shear $\Delta W / L = 0.40$, and the pre-compression is set as: (A) $\Delta L /L = 0.40$ (with stopping angle $\alpha_{\mathrm{stop}} = 0.36$) and (B) $\Delta L /L = 0.35$ (with $\alpha_{\mathrm{stop}} = 0.27$). The phases of $z(1/2) > 0$ (green circles) and $z(1/2) < 0$ (red squares) are separated by the solid line.}
\label{fig:modePhasePlot}
\end{figure}

It is noted that the mode selection of elastic ribbons under dynamic loading is governed by both geometric and materials' parameters. Figure~\ref{fig:modePhasePlot} shows the 2D phase diagrams in the $\{\dot{\alpha} - \nu \}$ space for two sets of parameters: $\Delta L / L = 0.40, \Delta W / L = 0.40, \alpha_{\mathrm{stop}} = 0.36$ for Fig.~\ref{fig:modePhasePlot}A; and $\Delta L / L = 0.35, \Delta W / L = 0.40, \alpha_{\mathrm{stop}} = 0.27$ for Fig.~\ref{fig:modePhasePlot}B. It shows that the mode selection is more sensitive to the loading rate for larger pre-compression ($\Delta L / L = 0.40$, Fig.~\ref{fig:modePhasePlot}A), but more sensitive to damping for smaller pre-compression ($\Delta L / L = 0.35$, Fig.~\ref{fig:modePhasePlot}B).

\section{Conclusion}
\label{sec:ConclusionSec}

In this study, we examined the dynamic snap-through instabilities of pre-deformed elastic ribbons under dynamic rotational boundary conditions. Our approach combined theoretical modeling, discrete numerical simulations, and tabletop experiments. The results show two distinct snap-through transition paths jumping through four stable equilibrium configurations (modes). The first path involves a single snap-through transition, similar to that of a planar beam. The second path features two nontrivial consecutive transitions that has not been reported previously. Path 2 brings a wealth of opportunities for mode skipping and selection.

We generated static bifurcation diagrams using both theoretical methods by solving the Kirchhoff equations with AUTO and discrete numerical simulations, finding good agreement between them. Under dynamic loading, our results  demonstrate delayed bifurcation in the snap-through process, revealing scaling trends in both time and angle. Since all snap-through behaviors here are determined by the saddle-node bifurcation, our results show that the delay scaling in the snap-through in ribbons mirrors that in beams. We quantified these scaling powers using asymptotic analysis of the planar beam equation.

The insights into the  delayed bifurcation of elastic ribbons allow us to manipulate the snap-through transition paths by skipping particular modes. By precisely controlling dynamic loading, we can skip specific modes, driven by either the bifurcation delay or the inertia effect. Moreover, dynamic loading conditions can be adjusted for mode selection. Using our efficient, robust numerical tool, we carried out extensive parametric studies, producing detailed phase diagrams for both mode skipping and mode selection. This work lays the groundwork for future dynamic snap-through studies in thin elastic structures and offers design principles for intelligent mechanical systems like metamaterials, precision instruments, and soft robotics.

\section*{ACKNOWLEDGMENTS}

K.J.H. acknowledges a research start-up grant (002271-00001) from the Nanyang Technological University, and financial support by the Ministry of Education (MOE) of Singapore under Academic Research Fund Tier 2 (T2EP50122-0001). M.L. acknowledges the Presidential Postdoctoral Fellowship from Nanyang Technological University, Singapore, and the start-up funding from the University of Birmingham, UK.

\appendix

\section{Symmetry of pre-deformed ribbons}
\label{sec:AppendixA}

\begin{figure}[t!]
\centering
\includegraphics[width=0.90\textwidth]{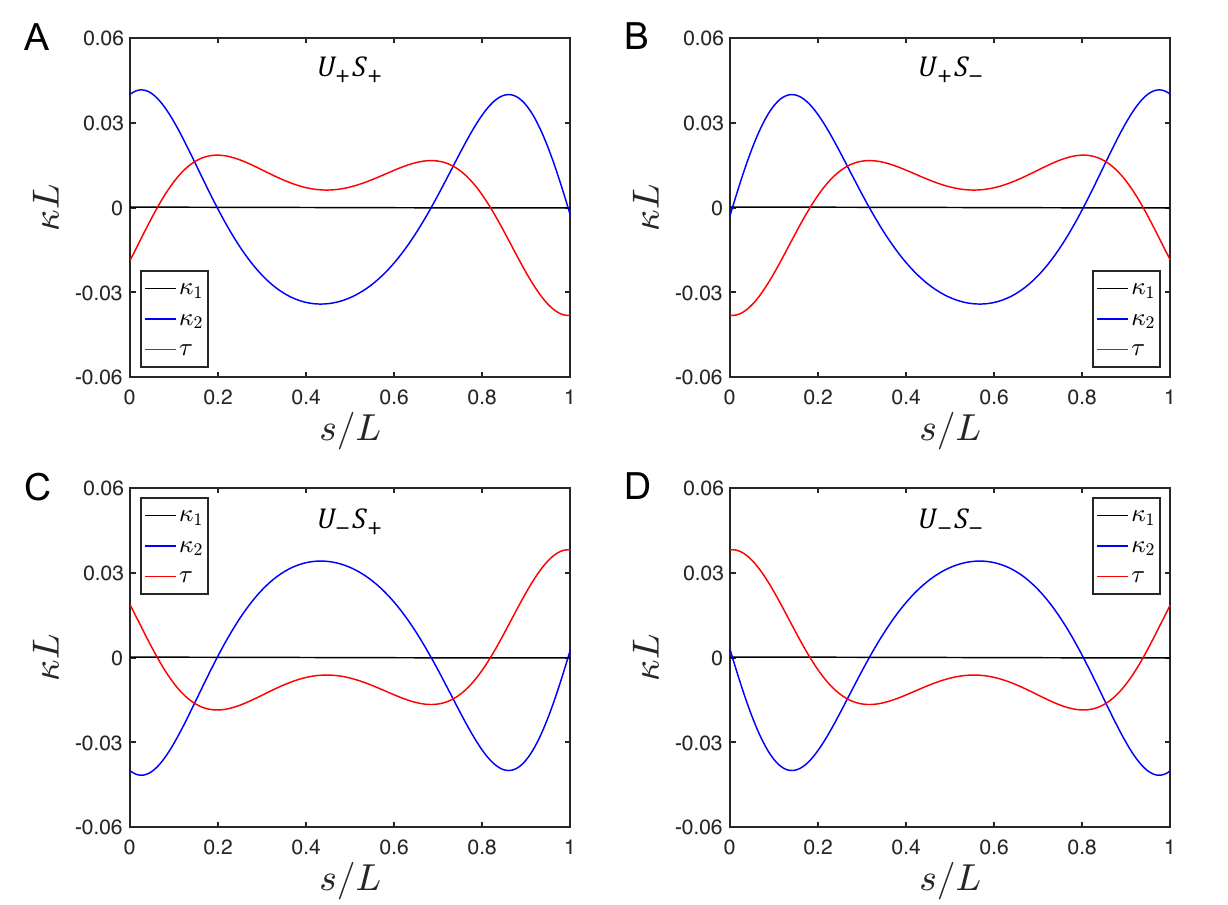}
\caption{Curvature distributions for an elastic ribbon with $\Delta L / L = 0.4$ and $\Delta W / L = 0.37$. (A) $U_+S_+$ (B) $U_+S_-$ (C) $U_-S_+$ (D) $U_-S_-$.}
\label{fig:curvaturePlot}
\end{figure}

In this appendix, we discuss the symmetry of ribbon's four static equilibria.
When the uniaxial compression is $\Delta L / L = 0.4$ and the transverse shear is $\Delta W / L = 0.37$, there are four static equilibrium configurations, referring to Fig.~\ref{fig:setupPlot}D, and some of them are symmetric (e.g., $U_+S_+$ and $U_-S_-$) and some of them are reverse symmetry (e.g., $U_+S_+$ and $U_+S_-$).
For a detailed comparison, in Fig.~\ref{fig:curvaturePlot}, the curvature distributions for all four configurations are provided.
On the one side, the curvatures would change of sign when $U_+$ becomes $U_-$, which is known as mirror symmetry; on the other side, the curvature distribution would inverse if $S_+$ varied to $S_-$, which can be regarded as reverse symmetry.

\section{Ribbon phase diagram}
\label{sec:AppendixB}

\begin{figure}[t!]
\centering
\includegraphics[width=0.90\textwidth]{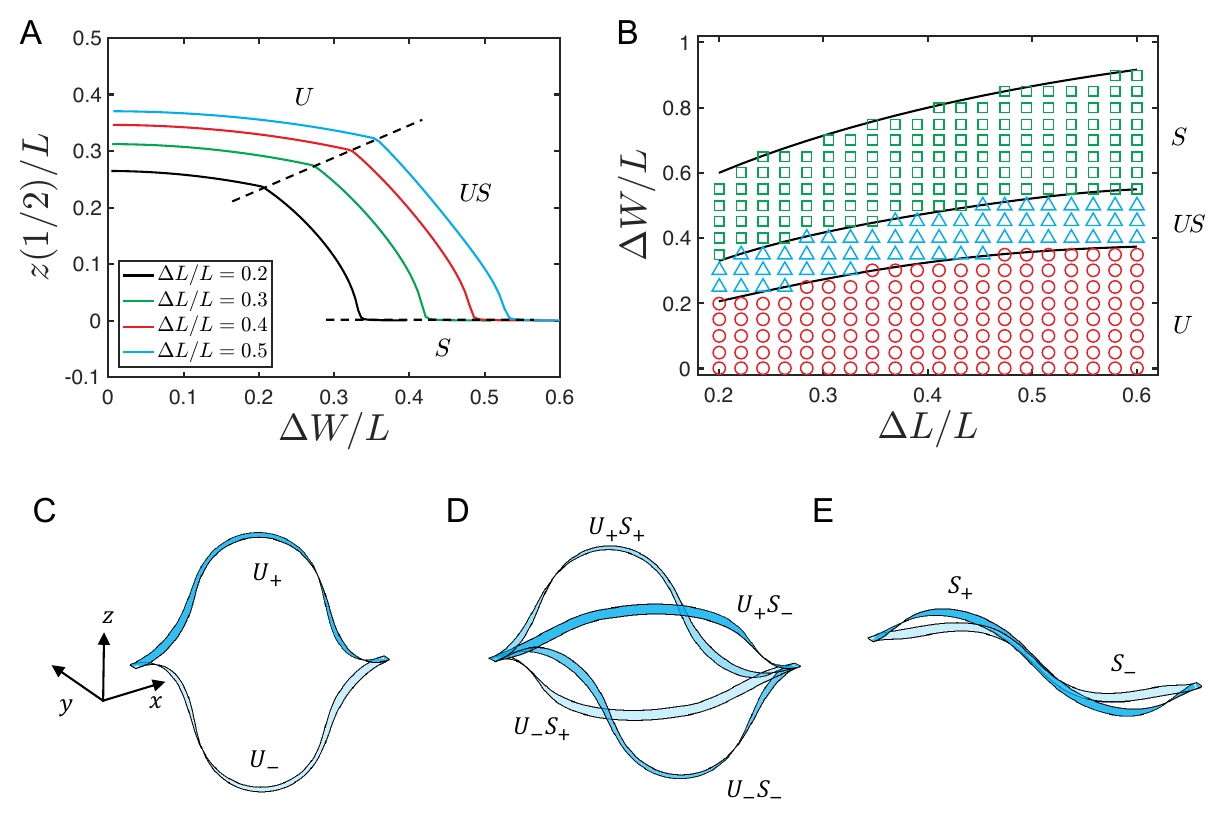}
\caption{(A) Normalized midpoint height, $z(1/2) / L$, as a function of transfer shear $\Delta W / L $, for different pre-compression, $ \Delta L / L \in \{0.2, 0.3, 0.4, 0.5\}$. (B) Phase diagram for a symmetric clamped-clamped ribbon, $U, US, S$. (C)-(E) Representative configurations for elastic ribbons.}
\label{fig:phaseApendixPlot}
\end{figure}

In this appendix, we review the configuration evaluation of a symmetric clamped-clamped ribbon with uniaxial compression $\Delta L$ and transverse shear $\Delta W$. More detailed analysis and results can be found in \cite{huang2021snap}.
In Fig.~\ref{fig:phaseApendixPlot}A, we plot the normalized midpoint height, $z(1/2) / L$, as a function of the transverse shear, $\Delta W / L$, for a pre-compressed ribbon (with $\Delta L / L \in \{ 0.20, 0.30, 0.40, 0.50 \}$).
Three different phases can be found in this bifurcation diagram, $U$, $US$, and $S$.
Next, a 2D parameter sweep, $\{\Delta L / L, \Delta W / L \}$, is performed to generate a phase diagram, as shown in Fig.~\ref{fig:phaseApendixPlot}B.
The associated configurations are available in Fig.~\ref{fig:phaseApendixPlot}C-E.
As expected, the shape transition between $U_+$ and $U_-$ would be similar to the classical 2D planar beam, while the snapping between $S_+$ and $S_-$ is impossible due to the inextensibility of the ribbon centerline.
In this work, we mainly focus on the shape morphing among $U_+S_+$, $U_+S_-$, $U_-S_+$, and $U_-S_-$.

\section{Video}
\label{sec:AppendixC}

We also provide four videos as supplementary material.
Movie-S1 is the mode $1$ snap-through process from desktop experiments, Movie-S2 is the mode $1$ snap-through process from our numerical simulation, Movie-S3 is the mode $2$ snap-through process from desktop experiments, and Movie-S4 is the mode $2$ snap-through process from our numerical simulation.

\bibliographystyle{elsarticle-harv}
\bibliography{dynamicRibbon}

\begin{thebibliography}{55}
\expandafter\ifx\csname natexlab\endcsname\relax\def\natexlab#1{#1}\fi
\expandafter\ifx\csname url\endcsname\relax
  \def\url#1{\texttt{#1}}\fi
\expandafter\ifx\csname urlprefix\endcsname\relax\def\urlprefix{URL }\fi

\bibitem[{Ameline et~al.(2017)Ameline, Haliyo, Huang, and
  Cognet}]{ameline2017classifications}
Ameline, O., Haliyo, S., Huang, X., Cognet, J.~A., 2017. Classifications of
  ideal 3{D} elastica shapes at equilibrium. Journal of Mathematical Physics
  58~(6).

\bibitem[{Antman and Jordan(1975)}]{antman19755}
Antman, S.~S., Jordan, K.~B., 1975. 5. — {Q}ualitative aspects of the spatial
  deformation of non-linearly elastic rods. {\S}. Proceedings of the Royal
  Society of Edinburgh Section A: Mathematics 73, 85--105.

\bibitem[{Audoly and Neukirch(2021)}]{audoly2021one}
Audoly, B., Neukirch, S., 2021. A one-dimensional model for elastic ribbons: A
  little stretching makes a big difference. Journal of the Mechanics and
  Physics of Solids 153, 104457.

\bibitem[{Audoly and Pomeau(2010)}]{audoly2010elasticity}
Audoly, B., Pomeau, Y., 2010. Elasticity and Geometry: From Hair Curls to the
  Non-linear Response of Shells. Oxford University Press.

\bibitem[{Bergou et~al.(2010)Bergou, Audoly, Vouga, Wardetzky, and
  Grinspun}]{bergou2010discrete}
Bergou, M., Audoly, B., Vouga, E., Wardetzky, M., Grinspun, E., 2010. Discrete
  viscous threads. In: ACM Transactions on Graphics (TOG). Vol.~29. ACM, p.
  116.

\bibitem[{Bergou et~al.(2008)Bergou, Wardetzky, Robinson, Audoly, and
  Grinspun}]{bergou2008discrete}
Bergou, M., Wardetzky, M., Robinson, S., Audoly, B., Grinspun, E., 2008.
  Discrete elastic rods. In: ACM transactions on graphics (TOG). Vol.~27. ACM,
  p.~63.

\bibitem[{Bertoldi et~al.(2017)Bertoldi, Vitelli, Christensen, and van
  Hecke}]{bertoldi2017flexible}
Bertoldi, K., Vitelli, V., Christensen, J., van Hecke, M., 2017. Flexible
  mechanical metamaterials. Nature Reviews Materials 2~(11), 17066.

\bibitem[{Bo et~al.(2023)Bo, Xu, Yang, and Zhang}]{bo2023mechanically}
Bo, R., Xu, S., Yang, Y., Zhang, Y., 2023. Mechanically-guided 3{D} assembly
  for architected flexible electronics. Chemical Reviews 123~(18),
  11137--11189.

\bibitem[{Brinkmeyer et~al.(2012)Brinkmeyer, Santer, Pirrera, and
  Weaver}]{brinkmeyer2012pseudo}
Brinkmeyer, A., Santer, M., Pirrera, A., Weaver, P., 2012. Pseudo-bistable
  self-actuated domes for morphing applications. International Journal of
  Solids and Structures 49~(9), 1077--1087.

\bibitem[{Camescasse et~al.(2013)Camescasse, Fernandes, and
  Pouget}]{camescasse2013bistable}
Camescasse, B., Fernandes, A., Pouget, J., 2013. Bistable buckled beam:
  {E}lastica modeling and analysis of static actuation. International Journal
  of Solids and Structures 50~(19), 2881--2893.

\bibitem[{Chen et~al.(2018)Chen, Bilal, Shea, and Daraio}]{chen2018harnessing}
Chen, T., Bilal, O.~R., Shea, K., Daraio, C., 2018. Harnessing bistability for
  directional propulsion of soft, untethered robots. Proceedings of the
  National Academy of Sciences 115~(22), 5698--5702.

\bibitem[{Chen et~al.(2023)Chen, Liu, and Jin}]{chen2023pseudo}
Chen, Y., Liu, T., Jin, L., 2023. Pseudo-bistability of viscoelastic shells.
  Philosophical Transactions of the Royal Society A 381~(2244), 20220026.

\bibitem[{Doedel(2007)}]{doedel2007lecture}
Doedel, E.~J., 2007. Lecture notes on numerical analysis of nonlinear
  equations. In: Numerical Continuation Methods for dynamical systems.
  Springer, pp. 1--49.

\bibitem[{Faris and Nayfeh(2007)}]{faris2007mechanical}
Faris, W., Nayfeh, A.~H., 2007. Mechanical response of a capacitive microsensor
  under thermal load. Communications in Nonlinear Science and Numerical
  Simulation 12~(5), 776--783.

\bibitem[{Forterre et~al.(2005)Forterre, Skotheim, Dumais, and
  Mahadevan}]{forterre2005venus}
Forterre, Y., Skotheim, J.~M., Dumais, J., Mahadevan, L., 2005. How the venus
  flytrap snaps. Nature 433~(7024), 421--425.

\bibitem[{Gomez et~al.(2017{\natexlab{a}})Gomez, Moulton, and
  Vella}]{gomez2017critical}
Gomez, M., Moulton, D.~E., Vella, D., 2017{\natexlab{a}}. Critical slowing down
  in purely elastic ‘snap-through’instabilities. Nature Physics 13~(2),
  142--145.

\bibitem[{Gomez et~al.(2017{\natexlab{b}})Gomez, Moulton, and
  Vella}]{gomez2017passive}
Gomez, M., Moulton, D.~E., Vella, D., 2017{\natexlab{b}}. Passive control of
  viscous flow via elastic snap-through. Physical Review Letters 119~(14),
  144502.

\bibitem[{Gomez et~al.(2019)Gomez, Moulton, and Vella}]{gomez2019dynamics}
Gomez, M., Moulton, D.~E., Vella, D., 2019. Dynamics of viscoelastic
  snap-through. Journal of the Mechanics and Physics of Solids 124, 781--813.

\bibitem[{Goyal et~al.(2005)Goyal, Perkins, and Lee}]{goyal2005nonlinear}
Goyal, S., Perkins, N.~C., Lee, C.~L., 2005. Nonlinear dynamics and loop
  formation in kirchhoff rods with implications to the mechanics of dna and
  cables. Journal of Computational Physics 209~(1), 371--389.

\bibitem[{Holmes and Crosby(2007)}]{holmes2007snapping}
Holmes, D.~P., Crosby, A.~J., 2007. Snapping surfaces. Advanced Materials
  19~(21), 3589--3593.

\bibitem[{Hu and Burgue{\~n}o(2015)}]{hu2015buckling}
Hu, N., Burgue{\~n}o, R., 2015. Buckling-induced smart applications: {R}ecent
  advances and trends. Smart Materials and Structures 24~(6), 063001.

\bibitem[{Huang et~al.(2021)Huang, Ma, and Qin}]{huang2021snap}
Huang, W., Ma, C., Qin, L., 2021. Snap-through behaviors of a pre-deformed
  ribbon under midpoint loadings. International Journal of Solids and
  Structures 232, 111184.

\bibitem[{Huang et~al.(2020)Huang, Wang, Li, and Jawed}]{huang2020shear}
Huang, W., Wang, Y., Li, X., Jawed, M.~K., 2020. Shear induced supercritical
  pitchfork bifurcation of pre-buckled bands, from narrow strips to wide
  plates. Journal of the Mechanics and Physics of Solids 145, 104168.

\bibitem[{Jawed et~al.(2018)Jawed, Novelia, and O'Reilly}]{jawed2018primer}
Jawed, M.~K., Novelia, A., O'Reilly, O.~M., 2018. A Primer on the Kinematics of
  Discrete Elastic Rods. Springer.

\bibitem[{Jiao and Liu(2020)}]{jiao2020snap}
Jiao, S., Liu, M., 2020. Snap-through in graphene nanochannels: With
  application to fluidic control. ACS Applied Materials \& Interfaces 13~(1),
  1158--1168.

\bibitem[{Keleshteri et~al.(2018)Keleshteri, Asadi, and
  Wang}]{keleshteri2018snap}
Keleshteri, M., Asadi, H., Wang, Q., 2018. On the snap-through instability of
  post-buckled {FG-CNTRC} rectangular plates with integrated piezoelectric
  layers. Computer Methods in Applied Mechanics and Engineering 331, 53--71.

\bibitem[{Liu et~al.(2023)Liu, Domino, de~Dinechin, Taffetani, and
  Vella}]{liu2023snap}
Liu, M., Domino, L., de~Dinechin, I.~D., Taffetani, M., Vella, D., 2023.
  Snap-induced morphing: From a single bistable shell to the origin of shape
  bifurcation in interacting shells. Journal of the Mechanics and Physics of
  Solids 170, 105116.

\bibitem[{Liu et~al.(2021)Liu, Gomez, and Vella}]{liu2021delayed}
Liu, M., Gomez, M., Vella, D., 2021. Delayed bifurcation in elastic
  snap-through instabilities. Journal of the Mechanics and Physics of Solids,
  104386.

\bibitem[{Liu et~al.(2019)Liu, Wu, Gan, Hanaor, and Chen}]{liu2019multiscale}
Liu, M., Wu, J., Gan, Y., Hanaor, D.~A., Chen, C., 2019. Multiscale modeling of
  the effective elastic properties of fluid-filled porous materials.
  International Journal of Solids and Structures 162, 36--44.

\bibitem[{Lu et~al.(2023{\natexlab{a}})Lu, Dai, Leanza, Hutchinson, and
  Zhao}]{lu2023multiple2}
Lu, L., Dai, J., Leanza, S., Hutchinson, J.~W., Zhao, R.~R.,
  2023{\natexlab{a}}. Multiple equilibrium states of a curved-sided hexagram:
  Part {II}—transitions between states. Journal of the Mechanics and Physics
  of Solids 180, 105407.

\bibitem[{Lu et~al.(2023{\natexlab{b}})Lu, Dai, Leanza, Zhao, and
  Hutchinson}]{lu2023multiple1}
Lu, L., Dai, J., Leanza, S., Zhao, R.~R., Hutchinson, J.~W.,
  2023{\natexlab{b}}. Multiple equilibrium states of a curved-sided hexagram:
  Part {I}—stability of states. Journal of the Mechanics and Physics of
  Solids 180, 105406.

\bibitem[{Ma et~al.(2022)Ma, Zhang, Jiao, and Liu}]{ma2022snap}
Ma, C., Zhang, Y., Jiao, S., Liu, M., 2022. Snap-through of graphene
  nanowrinkles under out-of-plane compression. Nanotechnology 34~(1), 015705.

\bibitem[{Mao et~al.(2022)Mao, Wang, Tan, and Liu}]{mao2022modular}
Mao, J.-J., Wang, S., Tan, W., Liu, M., 2022. Modular multistable metamaterials
  with reprogrammable mechanical properties. Engineering Structures 272,
  114976.

\bibitem[{Morigaki et~al.(2016)Morigaki, Wada, and
  Tanaka}]{morigaki2016stretching}
Morigaki, Y., Wada, H., Tanaka, Y., 2016. Stretching an elastic loop: Crease,
  helicoid, and pop out. Physical Review Letters 117~(19), 198003.

\bibitem[{Nizette and Goriely(1999)}]{nizette1999towards}
Nizette, M., Goriely, A., 1999. Towards a classification of
  {E}uler--{K}irchhoff filaments. Journal of Mathematical Physics 40~(6),
  2830--2866.

\bibitem[{Overvelde et~al.(2015)Overvelde, Kloek, D’haen, and
  Bertoldi}]{overvelde2015amplifying}
Overvelde, J.~T., Kloek, T., D’haen, J.~J., Bertoldi, K., 2015. Amplifying
  the response of soft actuators by harnessing snap-through instabilities.
  Proceedings of the National Academy of Sciences 112~(35), 10863--10868.

\bibitem[{Pandey et~al.(2014)Pandey, Moulton, Vella, and
  Holmes}]{pandey2014dynamics}
Pandey, A., Moulton, D.~E., Vella, D., Holmes, D.~P., 2014. Dynamics of
  snapping beams and jumping poppers. Europhysics Letters 105~(2), 24001.

\bibitem[{Qiao et~al.(2020)Qiao, Liu, and Pasini}]{qiao2020elastic}
Qiao, C., Liu, L., Pasini, D., 2020. Elastic thin shells with large
  axisymmetric imperfection: From bifurcation to snap-through buckling. Journal
  of the Mechanics and Physics of Solids 141, 103959.

\bibitem[{Qin et~al.(2023)Qin, Peng, Huang, Liu, and Huang}]{qin2023modeling}
Qin, L., Peng, H., Huang, X., Liu, M., Huang, W., 2023. Modeling and simulation
  of dynamics in soft robotics: A review of numerical approaches. Current
  Robotics Reports, 1--13.

\bibitem[{Radisson and Kanso(2023{\natexlab{a}})}]{radisson2023designing}
Radisson, B., Kanso, E., 2023{\natexlab{a}}. Designing shape transitions in
  elastic structures. Journal of the Physical Society of Japan 92~(12), 121010.

\bibitem[{Radisson and Kanso(2023{\natexlab{b}})}]{radisson2023dynamic}
Radisson, B., Kanso, E., 2023{\natexlab{b}}. Dynamic behavior of elastic strips
  near shape transitions. Physical Review E 107~(6), 065001.

\bibitem[{Rafsanjani et~al.(2015)Rafsanjani, Akbarzadeh, and
  Pasini}]{rafsanjani2016snapping}
Rafsanjani, A., Akbarzadeh, A., Pasini, D., 2015. Snapping mechanical
  metamaterials under tension. Advanced Materials 27~(39), 5931--5935.

\bibitem[{Reis et~al.(2018)Reis, Brau, and Damman}]{reis2018mechanics}
Reis, P.~M., Brau, F., Damman, P., 2018. The mechanics of slender structures.
  Nature Physics 14~(12), 1150--1151.

\bibitem[{Rodriguez et~al.(2023)Rodriguez, Calius, Khatibi, Orifici, and
  Das}]{rodriguez2023mechanical}
Rodriguez, S., Calius, E., Khatibi, A., Orifici, A., Das, R., 2023. Mechanical
  metamaterial systems as transformation mechanisms. Extreme Mechanics Letters,
  101985.

\bibitem[{Sano and Wada(2018)}]{sano2018snap}
Sano, T.~G., Wada, H., 2018. Snap-buckling in asymmetrically constrained
  elastic strips. Physical Review E 97~(1), 013002.

\bibitem[{Sano and Wada(2019)}]{sano2019twist}
Sano, T.~G., Wada, H., 2019. Twist-induced snapping in a bent elastic rod and
  ribbon. Physical Review Letters 122~(11), 114301.

\bibitem[{Simone(2007)}]{simone2007introduction}
Simone, A., 2007. An introduction to the analysis of slender structures.

\bibitem[{Thompson et~al.(1990)Thompson, Stewart, and
  Turner}]{thompson1990nonlinear}
Thompson, J. M.~T., Stewart, H.~B., Turner, R., 1990. Nonlinear dynamics and
  chaos. Computers in Physics 4~(5), 562--563.

\bibitem[{Urbach and Efrati(2020)}]{urbach2020predicting}
Urbach, E.~Y., Efrati, E., 2020. Predicting delayed instabilities in
  viscoelastic solids. Science Advances 6~(36), eabb2948.

\bibitem[{Wan et~al.(2024)Wan, Avis, Wang, Wang, Kusumaatmaja, and
  Zhang}]{wan2023finding}
Wan, G., Avis, S.~J., Wang, Z., Wang, X., Kusumaatmaja, H., Zhang, T., 2024.
  Finding transition state and minimum energy path of bistable elastic continua
  through energy landscape explorations. Journal of the Mechanics and Physics
  of Solids 183, 105503.

\bibitem[{Wang et~al.(2023)Wang, Wang, Liu, Qin, Cheng, Bolmin, Alleyne, Wissa,
  Baughman, Vella, et~al.}]{wang2023insect}
Wang, Y., Wang, Q., Liu, M., Qin, Y., Cheng, L., Bolmin, O., Alleyne, M.,
  Wissa, A., Baughman, R.~H., Vella, D., et~al., 2023. Insect-scale jumping
  robots enabled by a dynamic buckling cascade. Proceedings of the National
  Academy of Sciences 120~(5), e2210651120.

\bibitem[{Yang et~al.(2023)Yang, Zhou, Zhao, Huang, Wang, Hsia, and
  Liu}]{yang2023morphing}
Yang, X., Zhou, Y., Zhao, H., Huang, W., Wang, Y., Hsia, K.~J., Liu, M., 2023.
  Morphing matter: From mechanical principles to robotic applications. Soft
  Science 3~(4), 38.

\bibitem[{Yu and Hanna(2019)}]{yu2019bifurcations}
Yu, T., Hanna, J., 2019. Bifurcations of buckled, clamped anisotropic rods and
  thin bands under lateral end translations. Journal of the Mechanics and
  Physics of Solids 122, 657--685.

\bibitem[{Zhang et~al.(2019)Zhang, Ding, Qi, Tao, Wang, Zhao, Peng, and
  Shi}]{zhang2019multifunctional}
Zhang, Y., Ding, J., Qi, B., Tao, W., Wang, J., Zhao, C., Peng, H., Shi, J.,
  2019. Multifunctional fibers to shape future biomedical devices. Advanced
  Functional Materials 29~(34), 1902834.

\bibitem[{Zhang et~al.(2020)Zhang, Jiao, Wu, Ma, and
  Feng}]{zhang2020configurations}
Zhang, Y., Jiao, Y., Wu, J., Ma, Y., Feng, X., 2020. Configurations evolution
  of a buckled ribbon in response to out-of-plane loading. Extreme Mechanics
  Letters 34, 100604.

\end{thebibliography}




%% \bibitem must have one of the following forms:
%%   \bibitem[Jones et al.(1990)]{key}...
%%   \bibitem[Jones et al.(1990)Jones, Baker, and Williams]{key}...
%%   \bibitem[Jones et al., 1990]{key}...
%%   \bibitem[\protect\citeauthoryear{Jones, Baker, and Williams}{Jones
%%       et al.}{1990}]{key}...
%%   \bibitem[\protect\citeauthoryear{Jones et al.}{1990}]{key}...
%%   \bibitem[\protect\astroncite{Jones et al.}{1990}]{key}...
%%   \bibitem[\protect\citename{Jones et al., }1990]{key}...
%%   \harvarditem[Jones et al.]{Jones, Baker, and Williams}{1990}{key}...
%%


% \end{thebibliography}

\end{document}